\documentclass{article}

\usepackage{xcolor}
\definecolor{comment}{rgb}{0.0, 0.0, 0.0}
\definecolor{recent}{rgb}{0.0,0.0,0.0}
\definecolor{acc}{rgb}{0.0, 0.0, 0.0}
\usepackage{amsmath} 
\usepackage{amssymb}
\usepackage{textcomp}
\usepackage{bm}
\usepackage{graphicx}
\usepackage{graphics}
\usepackage{subcaption}
\usepackage[symbol]{footmisc}

\usepackage{PRIMEarxiv}

\usepackage[utf8]{inputenc} 
\usepackage[T1]{fontenc}    
\usepackage{hyperref}       
\usepackage{url}            
\usepackage{booktabs}       
\usepackage{amsfonts}       
\usepackage{nicefrac}       
\usepackage{microtype}      
\usepackage{lipsum}
\usepackage{fancyhdr}       
\usepackage{graphicx}       
\graphicspath{{media/}}     

\title{Exploring hierarchical framework of nonlinear sparse Bayesian learning algorithm
through numerical investigations
}

\author{
  Nastaran Dabiran \\
  Department of Civil and Environmental Engineering \\
  Carleton University  \\
  Ottawa, ON, Canada \\
   \\
   \And
  Brandon Robinson \\
  Department of Civil and Environmental Engineering \\
  Carleton University  \\
  Ottawa, ON, Canada \\
   \\
   \And
   Rimple Sandhu \\
   Computational Science Center \\
   National Renewable Energy Laboratory \\
    Golden, CO, United States\\
   \And
   Mohammad Khalil\thanks{\textit{Sandia National Laboratories is a multimission laboratory managed and operated by National Technology \& Engineering Solutions of Sandia, LLC, a wholly owned subsidiary of Honeywell International Inc., for the U.S. Department of Energy’s National Nuclear Security Administration under contract DE-NA0003525. This paper describes objective technical results and analysis. Any subjective views or opinions that might be expressed in the paper do not necessarily represent the views of the
U.S. Department of Energy or the United States Government.}} \\
   Quantitative Modeling \& Analysis Department \\
   Sandia National Laboratories\\
   Livermore, CA, United States \\
   \And
   Chris L. Pettit \\
   Aerospace Engineering Department \\
   US Naval Academy\\
   Annapolis, MD, United States\\
   \\
   \And
   Dominique Poirel \\
   Department of Mechanical and Aerospace Engineering\\
    Royal Military College of Canada\\  
    Kingston, ON, Canada\\
    \\
    \And
    Abhijit Sarkar\\
    Department of Civil and Environmental Engineering\\
    Carleton University\\
    Ottawa, ON, Canada\\
}

\begin{document}
\maketitle
\begin{abstract}

Sparse Bayesian learning (SBL) has been extensively utilized in data-driven modeling to combat the issue of overfitting. While SBL excels in linear-in-parameter models, its direct applicability is limited in models where observations possess nonlinear relationships with unknown parameters. Recently, a semi-analytical Bayesian framework known as nonlinear sparse Bayesian learning (NSBL) was introduced \textcolor{comment}{by the authors} to induce sparsity among model parameters during the Bayesian inversion of nonlinear-in-parameter models. NSBL relies on optimally selecting the hyperparameters of sparsity-inducing Gaussian priors. It is inherently an approximate method since the uncertainty in the hyperparameter posterior is disregarded as we instead seek the maximum \textit{a posteriori} (MAP) estimate of the hyperparameters (type-II MAP \textcolor{comment}{estimate)}. This paper aims to \textcolor{acc}{investigate the hierarchical structure that forms the basis of NSBL and} validate its accuracy through a comparison with \textcolor{comment}{a one-level} hierarchical Bayesian inference as a benchmark in the context of three numerical experiments: (i) a benchmark linear regression example with Gaussian prior and Gaussian likelihood, (ii) \textcolor{acc}{the same regression problem} with a highly non-Gaussian prior, and (iii) an example of a dynamical system with a non-Gaussian prior and a highly non-Gaussian likelihood function, to explore the performance of the algorithm in these new settings. Through these numerical examples, it can be shown that NSBL is \textcolor{comment}{well-suited} for physics-based models as it can be readily applied to models with non-Gaussian prior distributions and non-Gaussian likelihood functions. Moreover, we illustrate the accuracy of the NSBL algorithm as an approximation to \textcolor{comment}{the one-level} hierarchical Bayesian inference and its ability to \textcolor{acc}{reduce the computational cost} while adequately exploring the parameter posteriors.

\end{abstract}

\keywords{Bayesian inference, overfitting, sparse Bayesian learning, hierarchical Bayesian inference}

\section{Introduction}
\label{sec:intro} 

The Bayesian formalism provides a means to incorporate prior uncertainty and derive posterior uncertainty in the parameter inference of \textcolor{comment}{numerous} models \cite{bishop2006, hastie2009, murphy2012}. Various approaches exist within Bayesian inference, differing primarily in their assumptions regarding the prior distribution and how inference is performed \cite{murphy2012}. \textcolor{comment}{The one-level hierarchical Bayesian inference (from now on referred to as hierarchical Bayesian inference) }and sparse Bayesian learning (SBL) \cite{tipping1999, tipping2001, drugowitsch2013} are two prominent variants of Bayesian modeling that have gained significant attention in recent years. Hierarchical Bayesian inference involves the inclusion of multiple layers, allowing for complex dependencies among data points. However, Bayesian inference in high-dimensional parameter space quickly becomes intractable \textcolor{comment}{due to computational cost}; hence, the increase in dimensionality associated with the hierarchical Bayesian formulation exacerbates this issue. 
\textcolor{comment}{Several} efforts have focused on approximation techniques to infer the parameters and hyperparameters. \textcolor{comment}{In the empirical Bayesian framework,} Tipping \cite{tipping2001} proposed \textcolor{comment}{the} SBL, also known as \textcolor{comment}{the} relevance vector machine (RVM) \cite{faul2001, bishop2000}. This approach is effective in promoting sparsity \textcolor{comment}{in the parameter space (thereby reducing overfitting \cite{murphy2012})} and has been rapidly developed for the linear-in-parameter models \textcolor{acc}{and engineering mechanics applications \cite{huang2021sparse, filippitzis2022sparse, yuen2010bayesian}.}

SBL relies on the use of Gaussian automatic relevance determination (ARD) priors to enable the semi-analytical machinery that permits the iterative computation of Bayesian entities needed for sparse learning. This method is limited by its requirement that all priors be ARD priors and that the likelihood function be Gaussian. For weakly nonlinear problems, this can be alleviated by using the Laplace approximation for the likelihood function \cite{tipping2001}. However, in highly non-linear problems wherein the likelihood function is highly non-Gaussian, this approximation is no longer appropriate. Applying variational inference (VB), Tipping \cite{bishop2000} \textcolor{comment}{extended} RVM, known as variational RVM  \cite{drugowitsch2013}, whereby \textcolor{comment}{the approximate joint posterior probability density (pdf) of the parameters and hyperparameters are obtained}. While approximate, information regarding the uncertainty in the posterior distribution over both parameters and hyperparameters is available through this setup. However, VB is strict with the prior choice to satisfy its conjugacy \textcolor{comment}{(or semi-conjugacy)} requirements and, due to computational challenges, did not receive as much attention as \textcolor{comment}{previous approaches based on empirical Bayes}. \textcolor{acc}{Furthermore, for inference problems involving natural processes or engineered systems, many model parameters will have some prior information that modelers would like to incorporate through the use of informative parameter priors, which have not been considered in the aforementioned approaches.}

\textcolor{acc}{These practical issues are addressed by the semi-analytical framework known as nonlinear sparse Bayesian learning (NSBL) \cite{sandhu2020model,sandhu2021,sandhu2024encoding}. NSBL employs the use of a so-called hybrid prior, allowing a combination of Gaussian ARD priors and informative priors to be used. Critically, the semi-analytical machinery of the NSBL algorithm then relies on a Gaussian mixture model (GMM) approximation of the product of the likelihood function and the known informative prior. The existing body of works related to NSBL, has relied on this GMM approximation to derive expressions for the evidence and objective function for hyperparameter optimization. However, this crucial element of the algorithm has not been thoroughly validated. This paper aims to answer the following questions by examining the inner workings of the algorithm through a series of detailed numerical investigations: 
(i) can the GMM approximation affect the ultimate level of sparsity in the model?
(ii) is it possible for NSBL to yield a comparable solution to hierarchical Bayesian inference in situations consisting of non-Gaussian prior, non-Gaussian likelihood, or both? 
(iii) how does the reduced dimensionality of NSBL offer advantages over hierarchical Bayesian inference, and to what extent are the benefits of the additional level of hierarchy preserved? While prior NSBL-related works have been focused on practical aspects of NSBL, such as its proficiency as an alternative to model selection within a sparse learning framework \cite{sandhu2024encoding}, here, we offer a fresh perspective on NSBL by examining diverse scenarios involving Gaussian and non-Gaussian likelihood functions and priors. We examine the similarity between the hierarchical Bayesian inference and NSBL results considering aspects such as objective function/hyperparameter posterior as well as the suitability of the NSBL approximation for prior selection. Ultimately, by comparing the NSBL objective function with the hyperparameter posterior pdf derived from sampling in hierarchical Bayesian inference, we validate relevant approximations in numerous settings. We would contest that comparing the model forecasts obtained using NSBL and hierarchical Bayesian inference provides better support for the use of NSBL as a computationally efficient practical alternative to hierarchical Bayesian inference.}





For the sake of completeness, in Section \ref{sec.2}, we review the mathematical definition of hierarchical Bayesian inference and mathematical derivation of NSBL algorithm whereby the posterior \textcolor{comment}{is sampled using} transitional MCMC (TMCMC)~\cite{ching2007,betz2016}. In Section \ref{sec.3}, first, we consider a linear regression setting \textcolor{acc}{ with Gaussian prior and Gaussian likelihood which is then reused with a highly non-Gaussian prior to explore the performance of the algorithm in this new setting}. Next, we apply NSBL to a three-degree-of-freedom (dof) mass-spring-damper system where the nonlinearity arises in the likelihood function. 

\section{NSBL as an approximate hierarchical Bayesian setting}
\label{sec.2}
For a general nonlinear in parameter model of the form $f: \boldsymbol{\phi} \mapsto \mathbf{y}$, the unknown model parameter vector $\phi \in \mathbb{R}^{N_\phi}$ are mapped to the observed entity $\mathbf{y} \in \mathbb{R}^{N_y}$ with the model operator $f$. Given some (potentially noisy, sparse, and incomplete) training data $\mathcal{D}=\{\mathbf{x},\mathbf{y}\}$, the goal is \textcolor{comment}{to infer} parameter posterior,

\begin{equation} \label{2:posterior_standard}
\mathrm{p}(\bm{\phi} \vert \mathcal{D}) = \frac{\mathrm{p}(\mathcal{D}\vert\bm{\phi})\mathrm{p}(\bm{\phi})}{\mathrm{p}(\mathcal{D} ) }
=  \frac{\mathrm{p}(\mathcal{D}\vert\bm{\phi})\mathrm{p}(\bm{\phi})}{\int \mathrm{p}(\mathcal{D}\vert\bm{\phi})\mathrm{p}(\bm{\phi}) d\bm{\phi} }
\propto \mathrm{p}(\mathcal{D}\vert\bm{\phi})\mathrm{p}(\bm{\phi}),
\end{equation}
where the parameter prior $\mathrm{p}(\boldsymbol{\phi})$ is assigned based on some known information or is otherwise defined as non-informative, the likelihood function $\mathrm{p}(\mathcal{D}\vert\bm{\phi})$ is assumed to be known for every value of $\boldsymbol\phi$, and $\mathrm{p}(\mathcal{D})$ is the model evidence, \textcolor{comment}{which is an important quantity in the NSBL setting as will be explained later}.


In the following subsections, we delve into the discussion of hierarchical Bayesian setting and NSBL, along with their corresponding inference procedures.

\subsection{Hierarchical Bayesian Inference}
\label{sec.2.1}
Hierarchical Bayesian inference is a complex approach that captures multiple levels of uncertainty in the model parameters and is computationally \textcolor{comment}{intensive}. Assuming a \textcolor{comment}{one}-level hierarchy and applying Bayes' theorem, the joint posterior of the parameters $\boldsymbol{\phi}$ and hyperparameters $\boldsymbol{\alpha}$ is written as 

\begin{equation} \label{2:posterior}
\mathrm{p}(\bm{\phi},\bm{\alpha} \vert \mathcal{D}) = \frac{\mathrm{p}(\mathcal{D}\vert\bm{\phi})\mathrm{p}(\bm{\phi} \vert \bm{\alpha})\mathrm{p}(\bm{\alpha})}{\mathrm{p}(\mathcal{D}) },
\end{equation}
where the likelihood function $\mathrm{p}(\mathcal{D}\vert\bm{\phi})$ is assumed to be known for every value of $\boldsymbol\phi$. Note that the likelihood function is \textcolor{comment}{indirectly influenced by} the $\bm{\alpha}$ through the conditional dependence of the parameters $\bm{\phi}$ on the hyperparameters $\bm{\alpha}$. A key requirement for computing the posterior in this hierarchical Bayesian setup is the specification of the prior $\mathrm{p}(\boldsymbol{\phi} \mid \boldsymbol{\alpha})$ as will be discussed in Section \ref{sec2.1.2}. Finally, $\mathrm{p}(\mathcal{D})$ in the denominator denotes model evidence (or marginal likelihood or type-II likelihood). When the model evidence becomes too complex to calculate directly (i.e., involves a high-dimensional integral), one typically resorts to using sampling methods in order to obtain the posterior distribution of the model parameters \textcolor{comment}{\cite{sandhu2021,chib2001marginal}}. A well-known extension to classical MCMC that is applicable to complex distributions such as multimodal posterior pdfs and is capable of estimating model evidence directly is known as TMCMC \cite{ching2007}. 

\subsubsection{Prior and hyperprior pdf}
\label{sec2.1.2}
The Gaussian ARD prior of the form $\mathrm{p}\left(\boldsymbol{\phi} \mid \boldsymbol{\alpha}\right)=\mathcal{N}\left(\boldsymbol{\phi} \mid \mathbf{0}, \mathbf{A}^{-1}\right)$ is a popular type of prior distribution used in hierarchical Bayesian setting. The prior precision matrix (inverse covariance matrix) may be fully populated; however, assuming prior independence among the parameters leads to $\mathbf{A}=\operatorname{Diag}(\boldsymbol{\alpha})$. The key feature of the ARD prior is that they are zero mean distributions; thus, parameterizing the precision of the model parameters permits the relevance of each parameter to be estimated indirectly from the data
\textcolor{comment}{through data optimal prior precisions. While the proposed methodology is not restricted to the choice of specific hyperparameter prior, we use} Gamma distribution. \textcolor{comment}{This choice of hperprameter prior} enforces the requirement that the precision parameters $\boldsymbol{\alpha}$ be positive. Due to independence, the joint hyperprior $\mathrm{p}(\boldsymbol{\alpha})$ becomes as \cite{sandhu2021, sandhu2022}
\begin{equation} \label{2:gamma}
\begin{aligned}
\mathrm{p}(\boldsymbol{\alpha})=\prod_{i=1}^{N_\alpha} \mathrm{p}\left(\alpha_i\right)=\prod_{i=1}^{N_\alpha} \operatorname{Gamma}\left(\alpha_i \mid s_i, r_i\right)=\prod_{i=1}^{N_\alpha} \frac{r_i^{s_i}}{\Gamma\left(s_i\right)} \alpha_i^{s_i-1} e^{-r_i \alpha_i}, \quad   (s_i, r_i)>0,
\end{aligned}
\end{equation}
where the shape parameter $s_i$ and rate parameter $r_i$ are known parameters and $\operatorname{Gamma}\left(\alpha_i \mid s_i, r_i\right)$ denotes a univariate Gamma distribution. Although by varying $s_i$ and $r_i$, the Gamma distribution allows for the introduction of numerous simplified informative or non-informative distributions, the limit case of $s_i \rightarrow 0$ and $r_i \rightarrow 0$ is of particular interest in this paper. Setting these parameters to zero leads to a Jeffreys prior $p(\alpha_i) \propto 1/\alpha_i$ \textcolor{comment}{equivalently $p(\log\alpha_i) \propto 1$} for the hyperparameters. Jeffreys prior \cite{jeffreys1946} is a noninformative prior \textcolor{comment}{which} exhibits flatness over $\log \alpha_i$.
 The hyperprior in Eq.~(\ref{2:gamma}) transforms in log space \cite{papoulis2002,sandhu2022} as
\begin{equation} \label{2:logaplpha}
\begin{aligned}
\mathrm{p}(\log \boldsymbol{\alpha})=\prod_{i=1}^{N_\alpha} \mathrm{p}\left(\log \alpha_i\right)=\prod_{i=1}^{N_\alpha} \frac{\mathrm{p}\left(\alpha_i\right)}{\left|\frac{d}{d \alpha_i} \log \alpha_i\right|}=\prod_{i=1}^{N_\alpha} \frac{r_i^{s_i}}{\Gamma\left(s_i\right)} \alpha_i^{s_i} e^{-r_i \alpha_i}.
\end{aligned}
\end{equation}

\subsubsection{Predictive distribution}
\label{sec2.1.3}
The \textcolor{comment}{one}-level hierarchy in Eq.~(\ref{2:posterior}) with the Gaussian ARD prior and the Gamma hyperprior from Eq.~(\ref{2:gamma}) can be presented as $\boldsymbol{\alpha} \xrightarrow{}\boldsymbol{\phi} \xrightarrow{}\mathcal{D} $. Given the posterior distribution, we are interested in making predictions for a target $y^{*}$ using the predictive distribution \textcolor{comment}{{\cite{murphy2012}}}

\begin{equation} \label{2:prediction}
\begin{aligned}
\mathrm{p}\left(y^{*} \mid x^{*}, \mathcal{D}\right)=\int \mathrm{p}\left(y^{*} \mid  x^{*}, \boldsymbol{\phi}\right) \mathrm{p}( \boldsymbol{\phi}, \boldsymbol{\alpha} \mid \mathcal{D})  d \boldsymbol{\phi}d \boldsymbol{\alpha} ,
\end{aligned}
\end{equation}
marginalizing over the parameters and hyperparameters in the posterior pdf.

\subsection{Nonlinear sparse Bayesian learning}
\label{sec.2.2}
{\color{recent}
Typically, for the purpose of computing the parameter posterior, the model evidence is simply a normalization constant. However, for the purpose of sparse learning, and in Bayesian model selection in general, the model evidence is a critical quantity of interest \textcolor{comment}{\cite{murphy2012, bishop2006}}. It is an essential component of the objective function we define for the type-II MAP estimation of $\boldsymbol{\alpha}$. Within the sparse learning framework, the posterior distribution of the parameters $\boldsymbol{\phi}$, is conditional on the hyperparameters $\boldsymbol{\alpha}$, \textcolor{comment}{\cite{sandhu2020model, sandhu2021}}
  \begin{equation} \label{2:inference}
 \begin{aligned}
 \mathrm{p}(\boldsymbol{\phi} \mid \mathcal{D}, \boldsymbol{\alpha})=\frac{\mathrm{p}(\mathcal{D} \mid \boldsymbol{\phi}) \mathrm{p}(\boldsymbol{\phi} \mid \boldsymbol{\alpha})}{\mathrm{p}(\mathcal{D} \mid \boldsymbol{\alpha})}=\frac{\mathrm{p}(\mathcal{D} \mid \boldsymbol{\phi}) \mathrm{p}(\boldsymbol{\phi} \mid \boldsymbol{\alpha})}{\int \mathrm{p}(\mathcal{D} \mid \boldsymbol{\phi}) \mathrm{p}(\boldsymbol{\phi} \mid \boldsymbol{\alpha}) d \boldsymbol{\phi}}.
 \end{aligned}
 \end{equation}
whereas the parameters $\bm{\phi}$ and hyperparameters $\bm{\alpha}$ are jointly estimated in Eq.~(\ref{2:posterior}).

SBL (or RVM) is a widely used method for inducing sparsity in linear-in-parameter models of the form $\mathbf{y} = \boldsymbol{\psi}\boldsymbol{\phi} + \boldsymbol{\epsilon}$ such as regression problems \cite{tipping1999,tipping2001}, where $\boldsymbol{\psi}$ is the design matrix and $\boldsymbol{\epsilon}$ is the Gaussian model error with zero mean and precision $\rho$. However, it is limited by its requirement that all priors be ARD priors and that the likelihood function be Gaussian. The Gaussian ARD prior (as introduced in Section \ref{sec2.1.2}) and  the Gamma marginal hyperprior pdf $\mathrm{p}\left(\alpha_i\right)$ (introduced in Eq. (\ref{2:gamma})) are  popular choice  for the SBL setting.  This linearity property in the SBL setting and the Gaussian prior-posterior conjugacy offer a  semi-analytical  Bayesian analysis. As a result, the expression in Eq.~(\ref{2:inference}) is available analytically \cite{sandhu2021}. Note that the analytical tractability of SBL does not hold for nonlinear-in-parameter models or models with general non-Gaussian priors. In such instances, the semi-analytical framework known as nonlinear sparse Bayesian learning (NSBL) \cite{sandhu2020model,sandhu2021} addresses these practical issues through the use of a hybrid prior (in subsection \ref{sec2.2.1}) and a GMM approximation (in subsection \ref{sec2.2.2}).}


\subsubsection{Hybrid prior pdf}
\label{sec2.2.1}
In NSBL, the constraint on the choice of prior is relaxed by the adoption of a so-called hybrid prior. Following \cite{sandhu2017}, the concept of decomposing the set of parameters $\boldsymbol{\phi}= \{\boldsymbol{\phi}_\alpha, \boldsymbol{\phi}_{-\alpha}\}$ is implemented. Here, $\boldsymbol{\phi}_{-\alpha} \in \mathbb{R}^{N_\phi-N_\alpha}$ contains parameters that are \textit{a priori} relevant and have a known prior. The complementary set $\boldsymbol{\phi}_\alpha \in \mathbb{R}^{N_\alpha}$ is defined as the set of parameters whose relevance is \textit{a priori} unknown. Based on the sparsity inducing mechanism of SBL, $\boldsymbol{\phi_\alpha}$ is assumed to have a Gaussian ARD prior of the form $\mathrm{p}\left(\boldsymbol{\phi} \mid \boldsymbol{\alpha}\right)=\mathcal{N}\left(\boldsymbol{\phi} \mid \mathbf{0}, \mathbf{A}^{-1}\right)$. So, the joint prior pdf of $\boldsymbol{\phi}$ is denoted as
\begin{equation} \label{2:jointprior}
\begin{aligned}
\mathrm{p}(\boldsymbol{\phi} \mid \boldsymbol{\alpha}) = \mathrm{p}(\boldsymbol{\phi}_{-\alpha}) \mathrm{p} ({\boldsymbol\phi}_{\alpha}\mid \boldsymbol{\alpha}) = \mathrm{p}(\boldsymbol{\phi}_{-\alpha}) \mathcal{N}\left(\boldsymbol{\phi} \mid \mathbf{0}, \mathbf{A}^{-1}\right),
\end{aligned}
\end{equation}
Note that each parameter $\phi_i$ has a unique variable precision $\alpha_i$, such that we can write $\mathrm{p}\left(\phi_i \mid \alpha_i\right)=\mathcal{N}\left(\phi_i \mid 0, \alpha_i^{-1}\right)$. The hyperparameter, $\alpha_i$, dictates the complexity of the model by controlling the contribution of parameter $\phi_i$. Assigning ARD priors $\text{p}(\mathbf{\phi}_\alpha \mid \bm{\alpha})$ permits the automatic pruning of redundant parameters, while $\text{p}(\mathbf{\phi}_{-\alpha})$ encodes prior information about certain model parameters. 




The ARD prior $\mathrm{p} ({\boldsymbol\phi}_{\alpha}\mid \boldsymbol{\alpha})$ is conditioned on $\boldsymbol{\alpha}$ having a prior  $\mathrm{p}(\boldsymbol{\alpha})$. This hyperprior is relevant for the type-II MAP estimate of the hyperparameters, which involves the consideration of both the model evidence as well as the hyperprior. The Gamma marginal hyperprior pdf $\mathrm{p}\left(\alpha_i\right)$ is used as in  the hierarchical Bayesian setup  (see Section \ref{sec2.1.2}). 


\subsubsection{Gaussian mixture-model approximation}
\label{sec2.2.2}
The \textcolor{comment}{one}-level hierarchy in Eq.~(\ref{2:posterior}) is employed with the hybrid prior from Eq.~(\ref{2:jointprior}) and the Gamma hyperprior from Eq.~(\ref{2:gamma}). Constructing this hierarchical setting for sparse learning, the goal is defined as removing redundant model parameters and obtaining a sparse representation of unknown model parameter vector $\phi \in \mathbb{R}^{N_\phi}$. Given the hybrid prior defined in Eq.~(\ref{2:jointprior}), we rewrite Eq.~(\ref{2:inference}) as 

\begin{equation} \label{2:hybrid-posterior}
\mathrm{p}(\boldsymbol{\phi} \mid \mathcal{D}, \boldsymbol{\alpha})=\frac{\mathrm{p}(\mathcal{D} \mid \boldsymbol{\phi}) \mathrm{p}(\boldsymbol{\phi}_{-\alpha}) \mathcal{N}\left(\boldsymbol{\phi_{\alpha}} \mid \mathbf{0}, \mathbf{A}^{-1}\right)}{\mathrm{p}(\mathcal{D} \mid \boldsymbol{\alpha})}\propto \underbrace{\mathrm{p}(\mathcal{D} \mid \phi) \mathrm{p}\left(\phi_{-\alpha}\right)}_{\text {Independent of } \boldsymbol{\alpha}} \mathcal{N}\left(\phi_{\alpha} \mid \mathbf{0}, \mathbf{A}^{-1}\right).
\end{equation}

As noted, likelihood times the prior pdf of \textit{a priori} relevant parameters is an entity independent of $\boldsymbol{\alpha}$ parameter. Hence, there is no need to recalculate it as the algorithm iterates through different values of $\boldsymbol{\alpha}$ during optimization. For the sake of sparse learning, we construct a GMM of the form Eq.~(\ref{2:GMM}) \cite{sandhu2021,sandhu2022} 

\begin{equation} \label{2:GMM}
\mathrm{p}(\mathcal{D} \mid \phi) \mathrm{p}\left(\phi_{-\alpha}\right) \approx \sum_{k=1}^{K} a^{(k)} \mathcal{N}\left(\phi \mid \boldsymbol{\mu}^{(k)}, \boldsymbol{\Sigma}^{(k)}\right),
\end{equation}
where $K$ denotes the total number of kernels, $a^{(k)} \in \mathbb{R}$ is the kernel coefficient $(a^{(k)}>0, \sum_k^K a^{(k)} = 1)$ and $\left.\mathcal{N}(\boldsymbol{\phi} \mid \boldsymbol{\mu}^{(k)}, \boldsymbol{\Sigma}^{(k)})\right.$ is a Gaussian pdf with mean vector $\boldsymbol{\mu}^{(k)} \in \mathbb{R}^{N_\phi}$ and covariance matrix $\boldsymbol{\Sigma}^{(k)} \in \mathbb{R}^{N_\phi \times N_\phi}$ \cite{sandhu2021}. The use of Gaussian kernels
provides a semi-analytical Bayesian fremework (more details on mathematical derivation can be found in \cite{sandhu2020model, sandhu2021, sandhu2022}). Moreover, the use of a GMM relaxes the Gaussian assumptions in both the likelihood function and the known prior while retaining the analytical convenience of dealing with Gaussian distributions. The GMM can handle multimodal or skewed likelihood functions or a non-Gaussian prior pdf which we are dealing with in many engineering applications.

Note that in the case of SBL, due to the strict choice of ARD priors, this mixture modeling approximation is not necessary as everything is Gaussian, and the expression in Eq.~(\ref{2:GMM}) becomes exact with a single kernel.

\subsubsection{Sparse learning optimization problem}
\label{sec2.2.3}
Following SBL, for the sake of seeking sparsity in the set of questionable parameters $\{\boldsymbol{\phi}_\alpha\}$ our interest lies in the MAP estimate for the hyperparameter posterior $\mathrm{p}( \boldsymbol{\alpha} \mid \mathcal{D})$  \cite{tipping2001},

\begin{equation} \label{2:hyperparamposterior}
\begin{aligned}
\mathrm{p}( \boldsymbol{\alpha} \mid \mathcal{D})=  \mathrm{p}(\mathcal{D} \mid  \boldsymbol{\alpha}) \mathrm{p}( \boldsymbol{\alpha}),
\end{aligned}
\end{equation}
The first term $\mathrm{p}(\mathcal{D} \mid \boldsymbol{\alpha})$ is the model evidence in Eq.~(\ref{2:inference}) and $\mathrm{p}( \boldsymbol{\alpha})$ is the hyperprior pdf in Eq.~(\ref{2:gamma}). As we are interested in the type-II MAP estimate of the hyperparameters, the optimization of the hyperparameters can therefore be posed as \cite{murphy2012}
\begin{equation} \label{2:alphaMAP}
\begin{aligned}
\boldsymbol{\alpha}^{\text {MAP }}=\underset{\boldsymbol{\alpha}}{\arg \max }\{\mathrm{p}(\boldsymbol{\alpha} \mid \mathcal{D})\}=\underset{\boldsymbol{\alpha}}{\arg \max }\{\mathrm{p}(\mathcal{D} \mid \boldsymbol{\alpha}) \mathrm{p}(\boldsymbol{\alpha})\}.
\end{aligned}
\end{equation}

Alternatively, and more simply, we maximize the  $\log\mathrm{p}(\boldsymbol{\alpha} \mid \mathcal{D})$ \cite{sandhu2021}. Furthermore, if we prefer to perform optimization in terms of $\log \boldsymbol{\alpha}$, Eq.~(\ref{2:alphaMAP}), can be restated as


\begin{equation} \label{2:logalphaMAP}
\begin{aligned}
\log \boldsymbol{\alpha}^{\text {MAP }} 
& =\underset{\log \boldsymbol{\alpha}}{\arg \max }\{\log \mathrm{p}(\log \boldsymbol{\alpha} \mid \mathcal{D})\} ,\\
& =\underset{\log \boldsymbol{\alpha}}{\arg \max }\{\log \hat{\mathrm{p}}(\mathcal{D} \mid \log \boldsymbol{\alpha})+\sum_{i=1}^{N_\alpha}\log \mathrm{p}(\log \boldsymbol{\alpha}_i)\} ,
\end{aligned}
\end{equation}
whereby the terms independent of ${\alpha}_i$ are ignored, and the intractable model evidence is replaced by a GMM-based estimate of $\log \hat{\mathrm{p}}(\mathcal{D} \mid \log \boldsymbol{\alpha})$. Subsequently, the objective function $\mathcal{L}(\log \boldsymbol{\alpha})$ is derived by subtituting Eq.~(\ref{2:logaplpha}) in Eq.~(\ref{2:logalphaMAP}) \cite{sandhu2021,sandhu2022}, 

\begin{equation} \label{2:objective}
\begin{aligned}
\mathcal{L}(\log \boldsymbol{\alpha})=\log \hat{\mathrm{p}}(\mathcal{D} \mid \log \boldsymbol{\alpha})+\sum_{i=1}^{N_\alpha}\left(r_i \log \alpha_i-s_i \alpha_i\right).
\end{aligned}
\end{equation}

In SBL, this objective function can be obtained analytically, and the exact expression for the evidence may be used in place of the estimate shown here, 
\textcolor{recent}{
\begin{equation} \label{2:sbl_evidence}
\begin{aligned}
{\mathrm{p}}(\mathcal{D} \mid \boldsymbol{\alpha})= \int{\mathrm{p}(\mathcal{D}\vert \boldsymbol{\phi})\mathrm{p}(\boldsymbol{\phi}\vert \boldsymbol{\alpha})} = \mathcal{N}( \boldsymbol{0},\boldsymbol{\psi} \boldsymbol{A}^{-1}\boldsymbol{\psi} + \boldsymbol{I}\rho^{-1}),
\end{aligned}
\end{equation}} 
Consequently, the hyperparameter posterior $\mathrm{p}( \boldsymbol{\alpha} \mid \mathcal{D})$ in Eq. \ref{2:hyperparamposterior} is also available analytically and results in the analytical solution of $\boldsymbol{\alpha}^{\text {MAP }}$ in Eq. \ref{2:logalphaMAP}.  

Note that applying Jeffrey's prior ($s_i \approx 0$ and $r_i \approx 0$) as explained in Section \ref{sec2.1.2} results in reducing objective function to the first term (log-evidence) which is available in terms of the $K$ kernels of the GMM. Differentiating the objective function in Eq.~(\ref{2:objective}) with respect to $\log \alpha_i$ allows us to obtain expressions for the gradient vector. Differentiating once more provides the Hessian matrix, which permits the use of Newton's method for optimization, leveraging both the gradient and Hessian information \cite{sandhu2021,sandhu2020model}. Note that for SBL, this non-convex optimization becomes convex when optimizing the log-evidence with respect to each hyperparameter $\alpha_i$ individually \cite{sandhu2020model}. 

Once the hyperparameter MAP estimate $\log \alpha_i^{\text {MAP}}$ is determined, it is more reliable to use a scale-independent entity to identify relevant/irrelevant parameters. In SBL \cite{tipping2001}, the relevance indicator is defined as $\gamma_i=1-\alpha_i P_{i i}$, whereas in NSBL we define a similar metric for each individual kernel in the GMM approximation \textcolor{comment}{\cite{sandhu2020model,sandhu2021}}

\begin{equation}\label{2:relind}
\gamma_i^{(k)}=1-\frac{\alpha_i}{\left(P_{i i}^{(k)}\right)^{-1}} \in[0,1],
\end{equation}
and compute the root-mean-square value for each questionable parameter to provide a normalized metric on a scale of 0 to 1 \textcolor{comment}{indicating the} relevance \textcolor{comment}{of the} parameters. 




\section{Numerical investigations } 
\label{sec.3}
\begin{figure}[ht!]
\centering
\includegraphics[width=0.5\textwidth]{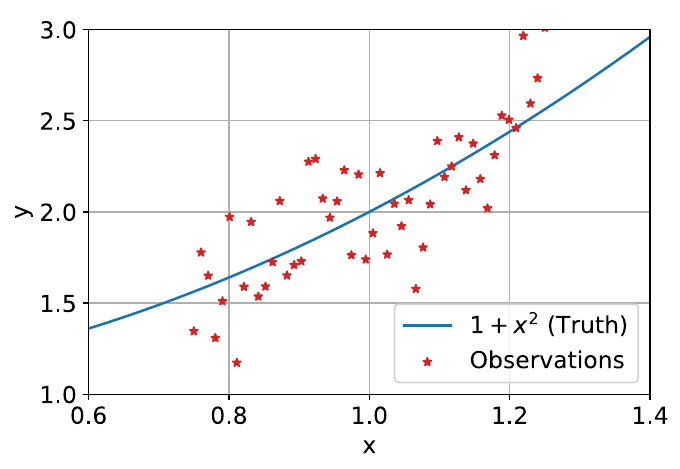}
\caption{Noisy observations versus the truth}
\label{fig:data}
\end{figure}

In this section, we perform \textcolor{recent}{three} numerical experiments i) Case 1 in subsection \ref{case1} with a Gaussian prior and Gaussian likelihood, ii) Case 2 in subsection \ref{sec.3.1} with non-Gaussian prior and Gaussian likelihood, and iii) Case 3 in subsection \ref{sec.3.2} with non-Gaussian prior and non-Gaussian likelihood. \textcolor{recent}{In Case 1, we use the analytical expressions of the parameter posterior pdf and model evidence available using SBL to validate the numerical implementation of hierarchical Bayesian inference using TMCMC in a linear regression setting. The next two investigations (Cases 2 and 3) are dedicated} to validating the accuracy of the NSBL algorithm vis-a-vis hierarchical Bayesian inference as a benchmark. In the second numerical experiment, we investigate how NSBL permits the data-optimal model discovery for a Bayesian linear regression exercise, wherein one of the parameters is assigned a highly-non Gaussian prior. In the third numerical experiment, we consider a more practical example of a multi-storey shear building frame where we only have prior knowledge of stiffness while damping is assumed questionable. In contrast to the second example, the third one emphasizes the critical importance of NSBL’s ability to function in the presence of highly non-Gaussian likelihood functions and non-Gaussian priors. The non-Gaussian nature of the likelihood function in this example is caused by the sparse temporal resolution of observations. 
\textcolor{recent}{\subsection{Case 1) Gaussian prior and Gaussian likelihood: Application of SBL in linear regression}}
In this section, we revisit the simple polynomial regression problem introduced by Sandhu et al. \cite{sandhu2021,sandhu2020model}, where we attempt to estimate the coefficients $a_0$, $a_1$, and $a_2$ of the second-order polynomial having the form

\begin{equation}
y = a_0 + a_1 x + a_2 x^2, \label{eq:polynomial}
\end{equation}
given some data generated according to the function 

\begin{equation}
y_i = 1 + x_i^2 + \epsilon_i, \quad \epsilon_i \sim \mathcal{N}(0,\rho^{-1}), \label{eq:poly_gen}
\end{equation}
where $\rho$ is the precision of the measurement noise. The data-generating model in Eq.~(\ref{eq:poly_gen}), can be recovered from Eq.~(\ref{eq:polynomial}) by assigning parameter values of $\phi = \{a_0 = 1, a_1 = 0, a_2 = 1\}$ and corrupting the observations by additive Gaussian noise. As shown in Figure \ref{fig:data}, the dataset consists of 50 evenly distributed points on $0.75 \leq x \leq 1.25$.
\textcolor{recent}{The parameter $a_0$, which represents the y-intercept, is assigned a Gaussian prior of the form $\text{p}(a_0) = \mathcal{N}(a_0 \vert 1,0.02^2)$. Given the ARD priors of the form $\mathrm{p}\left(\boldsymbol{\phi} \mid \boldsymbol{\alpha}\right)=\mathcal{N}\left(\boldsymbol{\phi} \mid \mathbf{0}, \mathbf{A}^{-1}\right)$ for the parameters $a_1$ and $a_2$, the joint posterior pdf of parameters $\boldsymbol{\phi}$ and hyperparameters $\boldsymbol{\alpha}$ can be expressed as Eq. (\ref{2:posterior})}. Note that the exact marginal posterior of $\boldsymbol{\alpha}$ is given in Eq. (\ref{2:sbl_evidence}).

\subsubsection{Sparse Bayesian Learning}\label{case1}
\begin{figure}[ht!]
\centering
\includegraphics[width=0.6\textwidth]{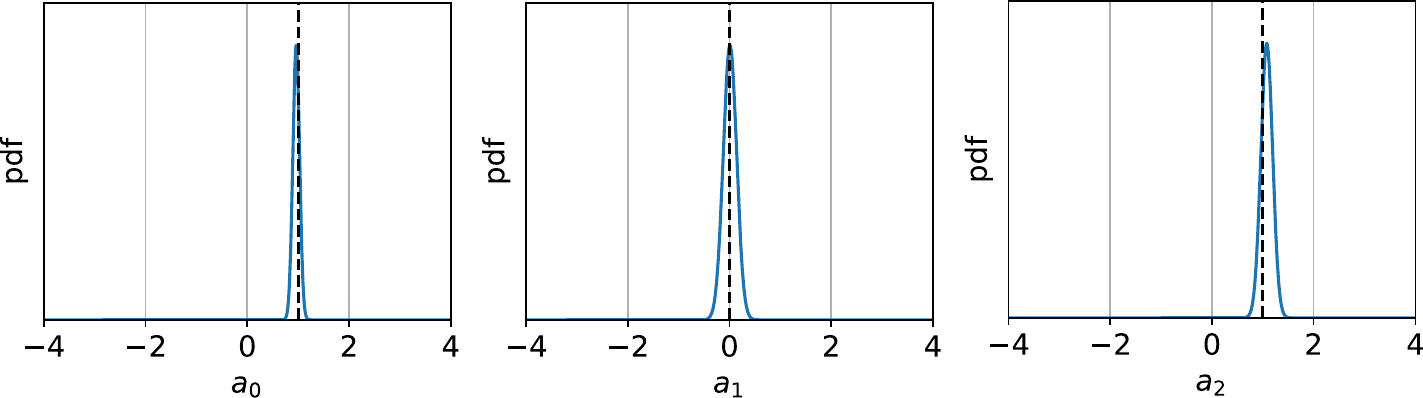}
\caption{Marginal posterior pdfs of parameters $a_0$, $a_1$, and $a_2$ obtained using SBL}
\label{fig:standard_post_params_1}
\end{figure}

For SBL, the problem can be stated based on Eq. \ref{2:inference} with the prior for a \textit{priori} relevant parameter given as a Gaussian and ARD prior assigned to questionable parameters as follow,

 \begin{equation} \label{3-1:inference}
\begin{aligned}
\mathrm{p}(\boldsymbol{\phi} \mid \mathcal{D}, \boldsymbol{\alpha})&=\frac{\mathrm{p}(\mathcal{D} \mid \boldsymbol{\phi}) \mathrm{p}(\boldsymbol{\phi} \mid \boldsymbol{\alpha})}{\mathrm{p}(\mathcal{D} \mid \boldsymbol{\alpha})}
\propto \mathcal{N}\left(\mathbf{y} \mid \bm{\phi}, {\rho}^{-1}\mathbf{I}\right) \mathcal{N}\left({a_0} \mid {1}, 0.02^2\right)\mathcal{N}\left({a_1} \mid {0}, \alpha_1^{-1}\right) \mathcal{N}\left({a_2} \mid {0}, \alpha_2^{-1}\right) 
\end{aligned}
\end{equation}
Considering the MAP estimation of $\alpha_1$ and $\alpha_2$, Eq. \ref{3-1:inference} is available analytically as stated in Eq. \ref{2:sbl_evidence} \cite{sandhu2021}. Consequently, the model evidence (the denominator in Eq.~(\ref{2:inference})) is also available analytically \cite{sandhu2021}. To sparsify $\phi$, we  find the mode or the MAP estimate of $\boldsymbol{\alpha}$ by maximizing the posterior of $\mathrm{p}(\boldsymbol{\alpha} \mid \mathcal{D})$ as explained in Section~\ref{sec2.2.3}. Upon convergence, the optimal hyperparameter values are obtained as $\boldsymbol{\alpha}^{MAP}= \{5.2, -0.81\}$.
Finally, the marginal posterior pdfs obtained using these optimal values are shown in Figure~\ref{fig:standard_post_params_1}. Clearly, the information gained from the data and informative Gaussian prior for parameter $a_0$ has resulted in the MAP of the posterior distribution being close to the actual parameter values.

\subsubsection{Hierarchical Bayesian inference}
\begin{figure}[ht!]
\centering
\includegraphics[width=0.6\textwidth]{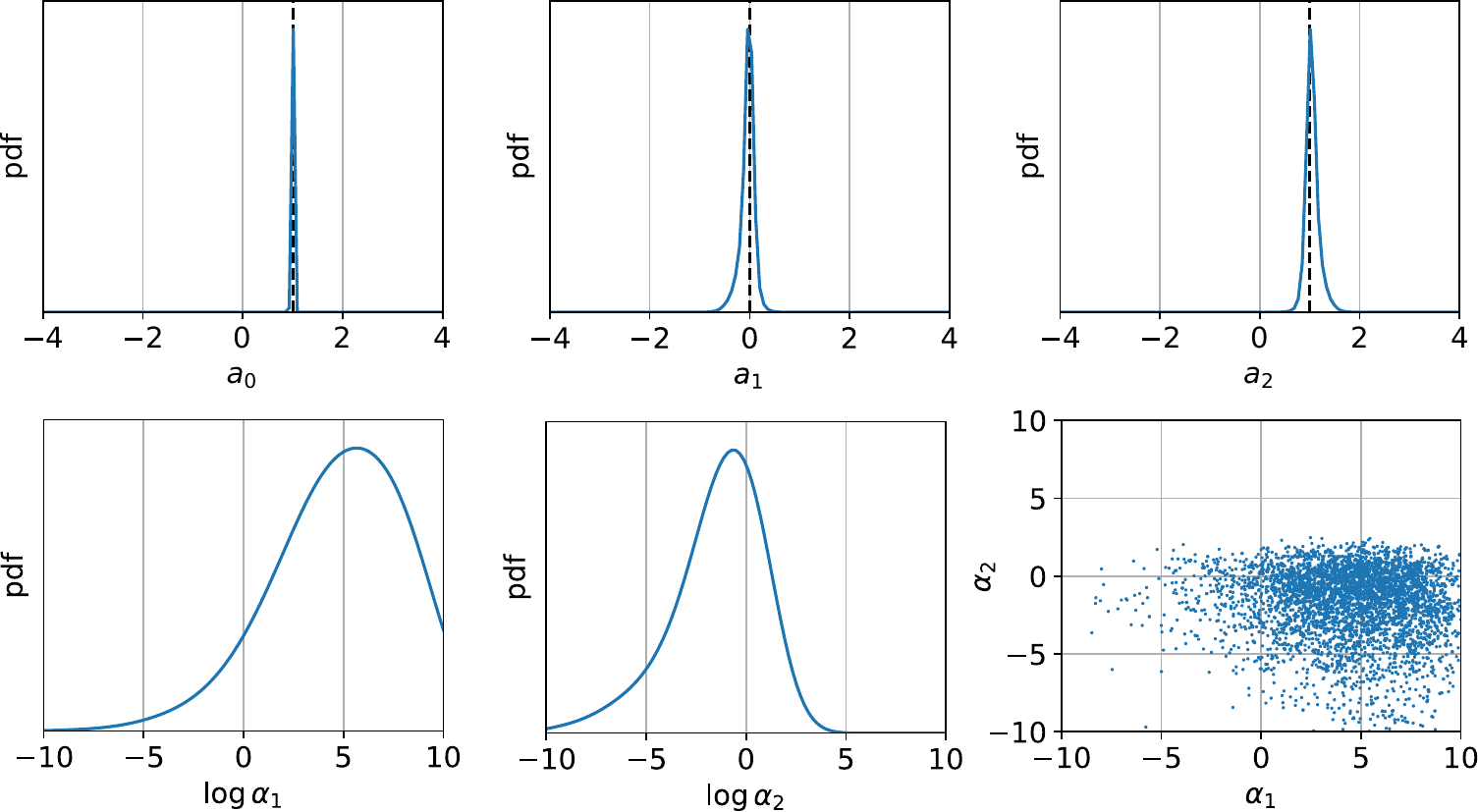}
\caption{Marginal posterior pdfs of parameters $a_0$, $a_1$, and $a_2$ and marginal posterior and joint samples of hyperparameters $\alpha_1$, and $\alpha_2$ obtained using hierarchical Bayesian inference}
\label{fig:hier_post_params_1}
\end{figure}

By employing ARD prior with precision $\alpha_1$ and $\alpha_2$ for the questionable parameters $a_1$ and $a_2$, and jointly estimating the parameters and hyperparameters ($a_0, a_1, a_2, \alpha_1, \alpha_2$) the inference problem can be stated as Eq. (\ref{2:inference}). The prior pdf of the \textit{a priori} relevant parameters $a_0$ is given by a Gaussian distribution of the form, and the prior pdfs of the potentially irrelevant parameters are assigned ARD prior. The hyperprior is given by Eq.~(\ref{2:gamma}) with shape and rate parameters $r_1 = r_2 = 1+\exp(-10)$ and $s_1 = s_2 = \exp(-10)$. This parameterization of the Gamma hyperprior results in an approximately uniform distribution in the range $\exp(-10) \leq \alpha_1, \alpha_2 \leq \exp(10)$. Thus, the hyperparameter posterior is largely data-driven, with the hyperprior providing an upper bound on the precision of a redundant parameter. The resulting parameter posterior pdfs and joint samples of hyperparameters $\alpha_1$, and $\alpha_2$ are shown in Figure \ref{fig:hier_post_params_1}.

This shows the automatic sparsity-inducing ability of hierarchical Bayesian inference for linear-in-parameter models. A sharp posterior prediction of parameter $a_1$ centered at zero is obtained, influenced by the unimodal hyperparameter posterior whose probability density is concentrated around large values of $\log \alpha_1$, resulting in a restrictive prior on parameter $a_1$.

\subsubsection{Comparisson of SBL and Hierarchical Bayesian inference }

\begin{figure}[ht!]
\begin{center}
\begin{subfigure}[b]{0.3\textwidth}
         \centering
         \includegraphics[width=\textwidth]{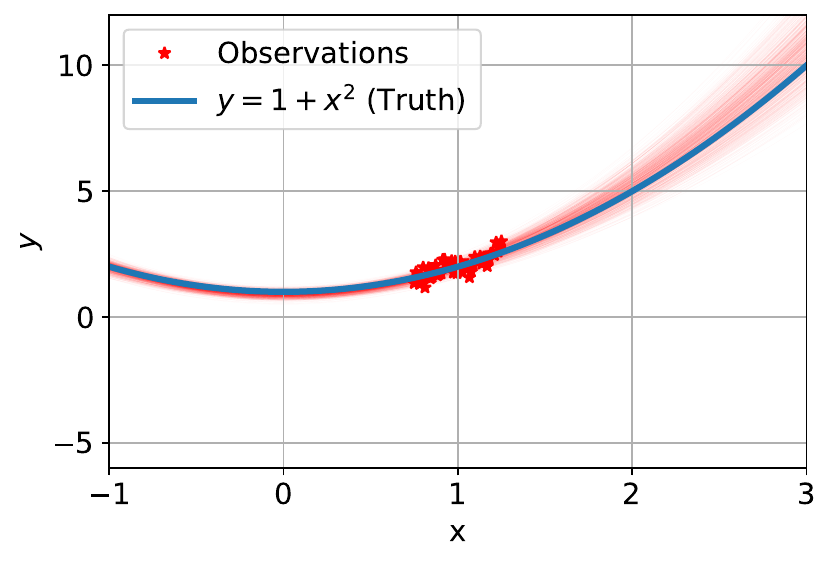}
         \caption{SBL}
         \label{fig:pred_nsbl}
     \end{subfigure}
\begin{subfigure}[b]{0.3\textwidth}
         \centering
         \includegraphics[width=\textwidth]{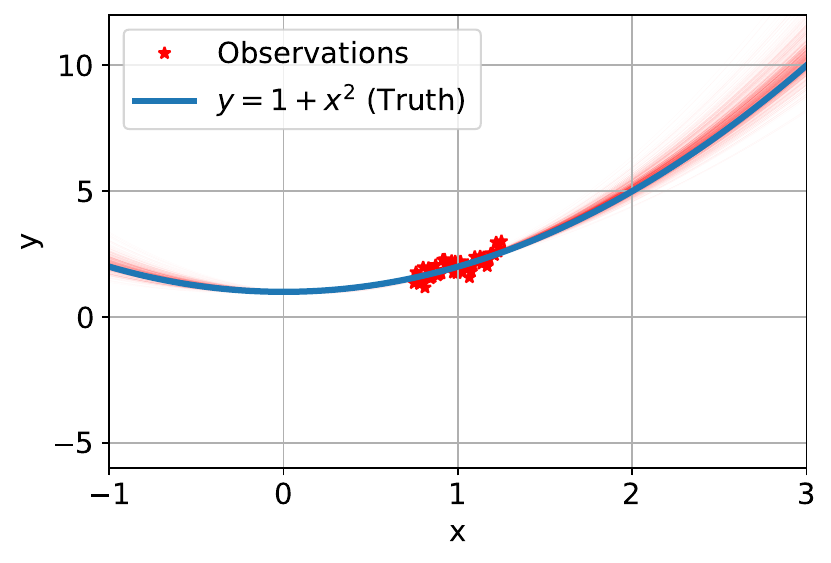}
         \caption{Hierarchical Bayesian inference}
         \label{fig:pred_hier}
     \end{subfigure}
\end{center}
\caption{Model predictions using 1000 samples from the parameter posterior pdfs. Panel (a) consists  of predictions made from joint samples of $\text{p}(a_0,a_1,a_2 \vert \mathcal{D},\alpha_1^\text{map},\alpha_2^\text{map})$. Panel (b) consists of predictions made from joint samples of $\text{p}(a_0,a_1,a_2,\alpha_1,\alpha_2 \vert \mathcal{D})$}
\label{fig:predictions_summary_1}
\end{figure}

In this section, we compare the objective function obtained using well-established SBL and hierarchical Bayesian inference. The resulting predictions are given in Figure \ref{fig:predictions_summary_1}. 

\begin{figure}[ht!]
\centering
\includegraphics[width=0.4\textwidth]{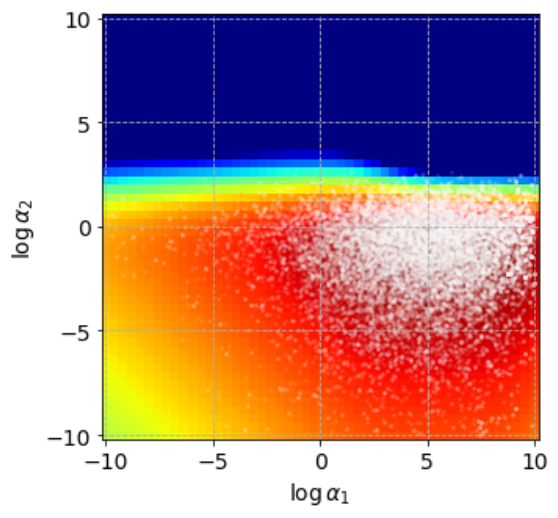}
\caption{Surface plots of the objective function for SBL superimposed by the joint samples from the hyperparameter posterior}
\label{fig:sbl_hierarchical}
\end{figure}

In Figure (\ref{fig:sbl_hierarchical}), the objective function obtained using SBL superimposed with the joint hyperparameter samples from hierarchical Bayesian inference. This figure highlights the close alignment between the exact analytical expression (SBL) and samples generated using TMCMC for hierarchical Bayesian inference, establishing confidence in our implementation.

\subsection{Case 2) Non-Gaussian prior and Gaussian likelihood: Application of NSBL in linear regression with a trimodal prior}
\label{sec.3.1}
\begin{figure}[ht!]
\begin{center}
         \centering
         \includegraphics[width=0.35\textwidth]{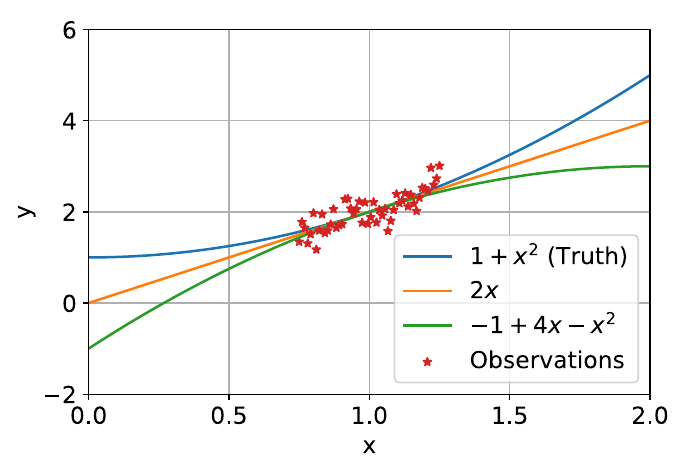}
\end{center}
\caption{Noisy observations and the three possible polynomial functions that fit the data and respect the prior on parameter $a_0$}
\label{fig:trimodal_data}
\end{figure}

In this section, we revisit the polynomial regression problem introduced by Sandhu et al. \cite{sandhu2021,sandhu2020model} (as explained in the first example), by assigning the parameter $a_0$, a multimodal prior of the form,


\begin{equation}
\text{p}(a_0) = \mathcal{N}(a_0 \vert -1,0.02^2) + \mathcal{N}(a_0 \vert 0,0.02^2) + \mathcal{N}(a_0 \vert 1,0.02^2), \label{eq:prior_a0}
\end{equation}
where each kernel has low variance, such that there is a region of low-probability between the distinct kernels. Thus, if we consider the parameter prior pdfs for $a_1$ and $a_2$ to be non-informative \textcolor{comment}{(uniform, $\text{p}(a_1)\propto 1$ , $\text{p}(a_2)\propto 1$)}, the parameter posterior pdf can be expressed as

\begin{equation}\label{eq:standard_bayes_noninformative}
\text{p}(\bm{\phi} \vert \mathcal{D}) = \frac{\text{p}(\mathcal{D}\vert\bm{\phi} )\text{p}(\bm{\phi})}{\text{p}(\mathcal{D} )} \propto \text{p}(\mathcal{D}\vert\bm{\phi} )\text{p}(a_0)\text{p}(a_1)\text{p}(a_2) \textcolor{comment}{\propto \text{p}(\mathcal{D}\vert\bm{\phi} )\text{p}(a_0)}
\end{equation}
where the evidence in the denominator and the uniform priors on $a_1$ and $a_2$ in the numerator are all constants.  \textcolor{comment}{Notice that this setting is the standard Bayesian method whereby the 
parameters with no prior knowledge are assigned non-informative prior.} The likelihood function requires that the parameters are selected such that the solution $y$ passes through the cluster of noisy data, while the prior requires that the solutions also pass through the y-intercept in the immediate vicinity of either $y = 1, 0, -1$. As shown in Figure~\ref{fig:trimodal_data}, this results in solutions in the neighbourhood of the following second-order polynomials:

\begin{subequations}
\begin{align}
        y & = 1 + x^2 && (a_0=1, a_1 = 0, a_2 = 1), \label{eq:poly_sol1} \\
        y & = 2x && (a_0=0, a_1 = 2, a_2 = 0), \label{eq:poly_sol2} \\ 
        y & = -1 + 4x - x^2 && (a_0=-1, a_1 = 4, a_2 = -1). \label{eq:poly_sol3}          
\end{align}
\end{subequations}


\begin{figure}[ht!]
\centering
\includegraphics[width=\textwidth]{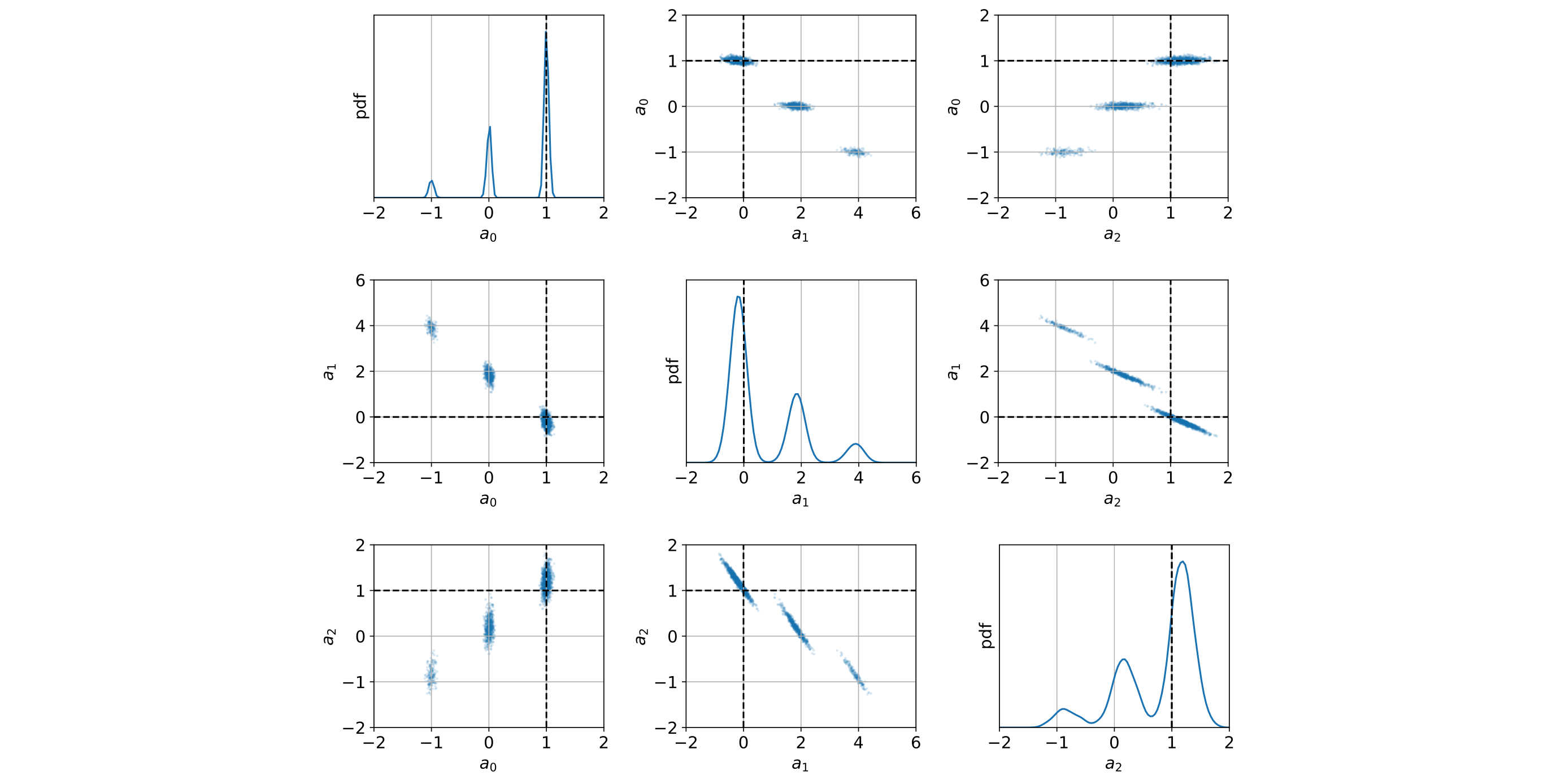}
\caption{Marginal and (pairwise) joint posterior pdfs of parameters $a_0$, $a_1$, and $a_2$ obtained using standard Bayesian inference. This is equivalent to the partial posterior given by the product of the known prior and the likelihood function.}
\label{fig:standard_post_params}
\end{figure}

The marginal posterior pdf and the pairwise-joint TMCMC samples of the parameter posterior are shown in Figure~\ref{fig:standard_post_params}. Each mode in the marginal plots, and each cluster of samples in the joint scatterplots correspond to one of the three combinations of parameters outlined in Eqs. (\ref{eq:poly_sol1})-(\ref{eq:poly_sol3}). Note the significant correlation between the linear coefficient $a_1$ and the quadratic coefficient $a_2$, evidenced through the joint samples.

The most prominent peak in the marginal pdfs, and the cluster of samples having the highest density in the joint pdfs in Figure~\ref{fig:predictions_summary} are centered at the \textit{true} parameter value  (identified by a dashed line). Clearly, the information gained from the data has resulted in the MAP of the posterior distribution being relatively close to the actual parameter values. However, due to the sparsity and noise of the data, the parameter posterior distributions remain multimodal due to the strong influence of the prior relative to that of the likelihood function (the predictive distribution is shown later in Figure~\ref{fig:pred_stand}.

\subsubsection{Nonlinear Sparse Bayesian Learning}
\begin{figure}[ht!]
\begin{center}
\begin{subfigure}[b]{0.3\textwidth}
         \centering
         \includegraphics[width=\textwidth]{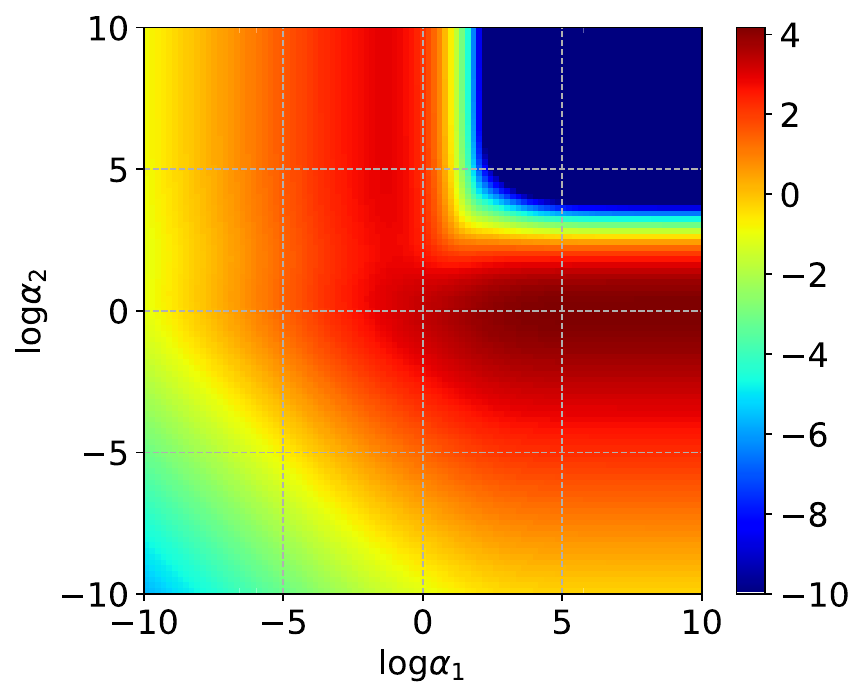}
         \caption{Evidence}
         \label{fig:nsbl_evidence}
     \end{subfigure}
\begin{subfigure}[b]{0.3\textwidth}
         \centering
         \includegraphics[width=\textwidth]{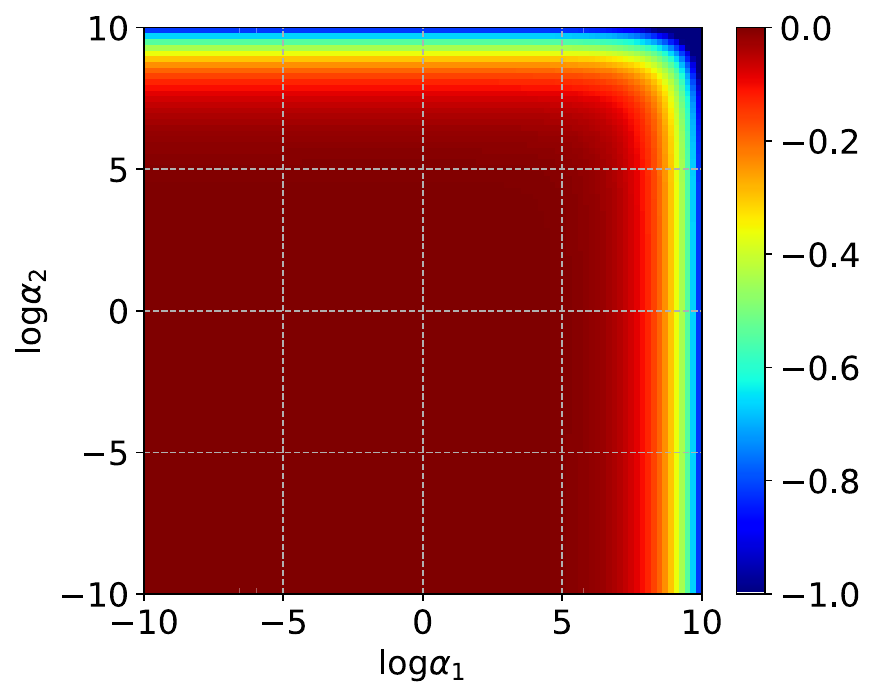}
         \caption{Hyperprior}
         \label{fig:nsbl_hyperprior}
     \end{subfigure}
\begin{subfigure}[b]{0.3\textwidth}
         \centering
         \includegraphics[width=\textwidth]{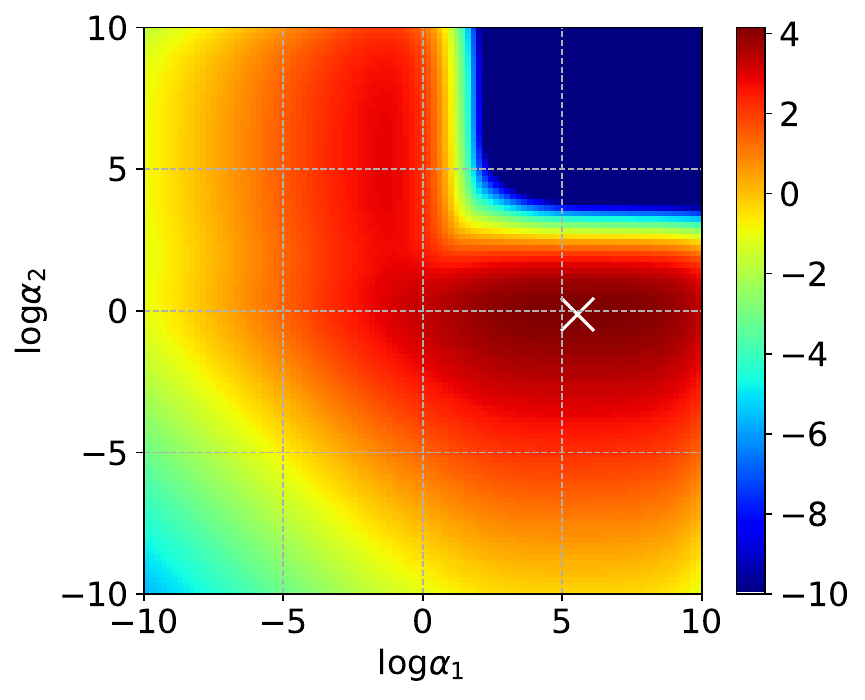}
         \caption{Objective function}
         \label{fig:nsbl_objective}
     \end{subfigure}
\end{center}
\caption{Surface plots of the model evidence, hyperprior and objective function for NSBL as a function of $\log\alpha_1$ and $\log\alpha_2$}
\label{fig:nsbl_surfs}
\end{figure}

\begin{figure}[ht!]
\begin{center}
\begin{subfigure}[b]{0.3\textwidth}
         \centering
         \includegraphics[width=\textwidth]{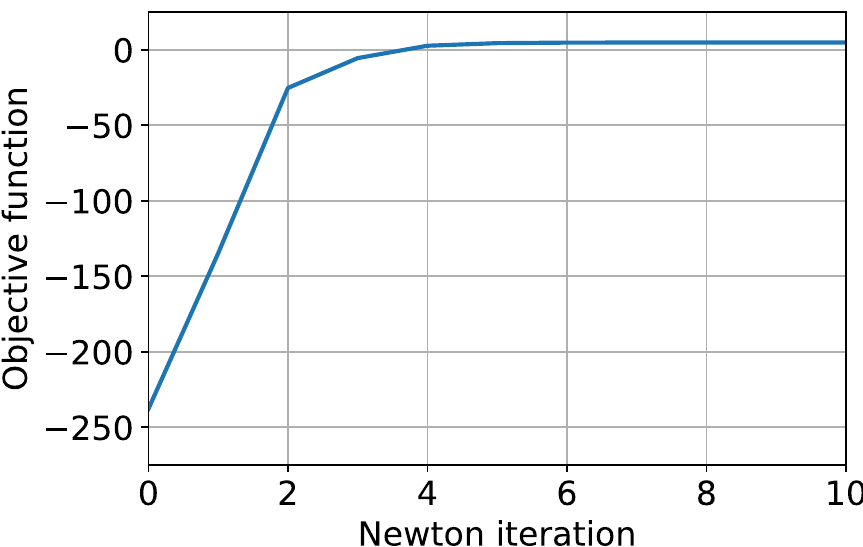}
         \caption{}
         \label{fig:poly_objfun}
     \end{subfigure}
\begin{subfigure}[b]{0.3\textwidth}
         \centering
         \includegraphics[width=\textwidth]{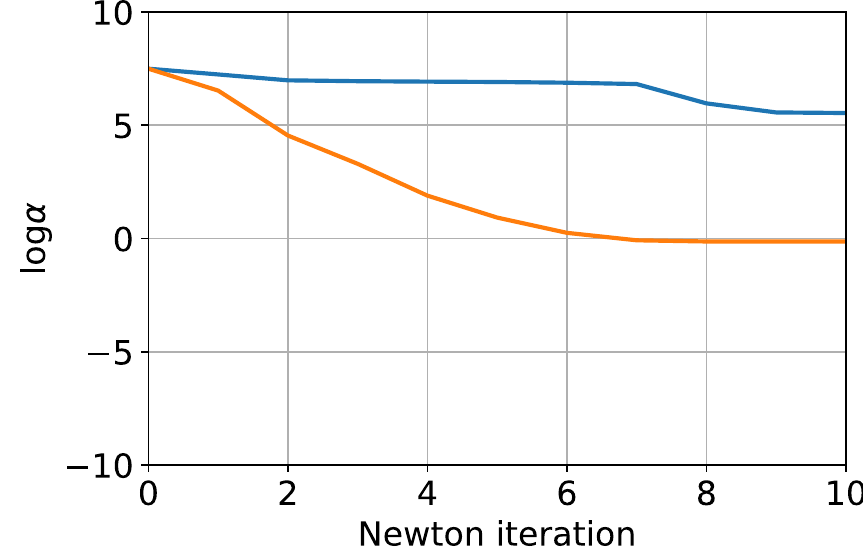}
         \caption{}
         \label{fig:poly_logalpha}
     \end{subfigure}
\begin{subfigure}[b]{0.3\textwidth}
         \centering
         \includegraphics[width=\textwidth]{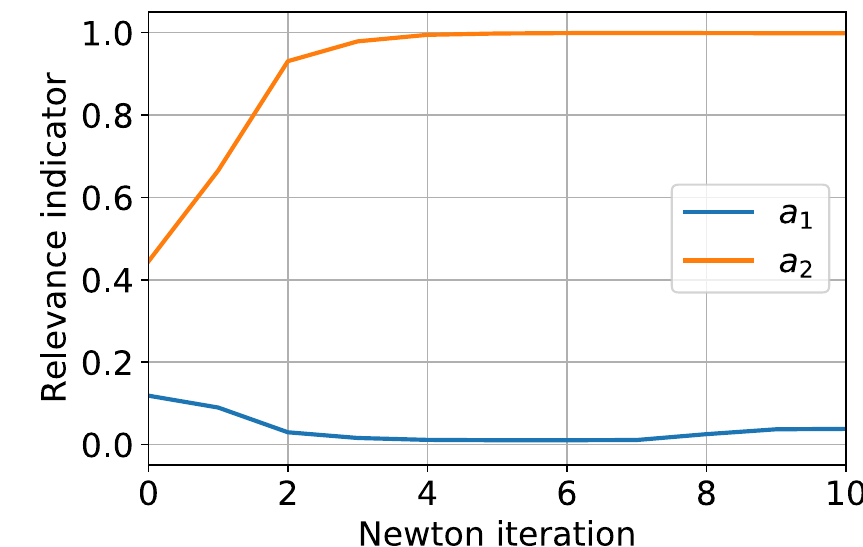}
         \caption{}
         \label{fig:poly_relind}
     \end{subfigure}
\end{center}
\caption{NSBL optimization as a function of Newton iterations}
\label{fig:nsbl_optimization}
\end{figure}

For NSBL, the inference problem can be stated as in Eq.~(\ref{2:posterior}), with the prior for the \textit{a priori} relevant parameters given by Eq.~(\ref{eq:prior_a0}), and an ARD prior assigned to the questionable parameters,

\begin{equation}
\text{p}(\bm{\phi}_\alpha \vert \bm{\alpha}) =  \mathcal{N}(a_1 \vert 0,\alpha_{1}^{-1}) \mathcal{N}(a_2\vert 0,\alpha_2^{-1}), \label{eq:prior_phi_alpha}.
\end{equation}

where the hyperprior is given by Eq.~(\ref{2:gamma}) with shape and rate parameters $\log r_i = \log s_i = -10$. Note that the sampling portion of the algorithm is only concerned with the product of the likelihood and the known prior as needed for the construction of the GMM in Eq.~(\ref{2:GMM}) \textcolor{comment}{ This is equivalent to the expression for the unnormalized posterior in Eq.~(\ref{eq:standard_bayes_noninformative}) with non-informative priors for the questionable parameters $a_1$ and $a_2$}. Kernel density estimation (KDE) has been used to construct a GMM of the likelihood times known prior in Figure~\ref{fig:standard_post_params}. \textcolor{comment}{Using this GMM approximation,} we can directly obtain estimates of the model evidence and objective function \textcolor{comment}{both available analytically \cite{sandhu2020model,sandhu2021,sandhu2022}} as a function of hyperparameters $\alpha_1$ and $\alpha_2$. In Figure~\ref{fig:nsbl_surfs}, we provide a visualization of the estimate of the model evidence, the hyperprior and the resulting objective function. From Figure \ref{fig:nsbl_evidence}, it can be seen that there is no clear optimum in the log evidence, but rather a large flat region along the axis $\log\alpha_2=0$, corresponding to a low-precision prior on $a_2$. Along the axis $\log\alpha_2=0$, the evidence function increases monotonically with increasing values of $\log\alpha_1$. In general, as the value of $\log\alpha_1$ approaches infinity, the relevance indicator will asymptotically approach zero. However, in practice, a finite value of $\log\alpha_1$ is sufficient to classify the parameter $a_1$ as irrelevant. For instance, from the two rightmost panels of Figure \ref{fig:nsbl_optimization}, it can be observed that a value of $\log\alpha_1 = 5.55$ gives a relevance indicator value of $\gamma_1^{rms} = 0.0375$. Also noteworthy is the upper right quadrant of Figure \ref{fig:nsbl_evidence} has significantly lower evidence than the rest of the domain. This region corresponds to high precision priors for both $a_1$ and $a_2$, implying both parameters would be redundant, resulting in a model of the form $y=a_0$. The log of the Gamma prior pdf on $\log\bm{\alpha}$ is depicted in Figure \ref{fig:nsbl_hyperprior}. The hyperprior  is parameterized such that $\log \text{p}(\log \bm{\alpha})$ is approximately flat over most of the domain, and begins to decrease exponentially in the immediate vicinity of the upper bounds of the domain. This produces the desirable effect of regularizing objective function, resulting in a unique optimum as seen in Figure \ref{fig:nsbl_objective} (identified by the $\times$). 

\textcolor{comment}{Figure \ref{fig:poly_objfun} shows the objective function against Newton's iteration}. The convergence to the optimum at $(\log\alpha_1 = 5.55, \log\alpha_2 = -0.136)$ approached from an initial coordinate in the low-evidence region $(\log\alpha_1 = 7.50, \log\alpha_2 = 7.50)$ is achieved in 10 iterations \textcolor{comment}{(see Figure \ref{fig:poly_logalpha})}. \textcolor{comment}{ In Figure \ref{fig:poly_relind}} the relevance indicator is given in Eq.~(\ref{2:relind}) converges to a value of 0.0375 for parameter $a_1$ and to a value of 0.999 for $a_2$.

\begin{figure}[ht!]
\centering
\includegraphics[width=\textwidth]{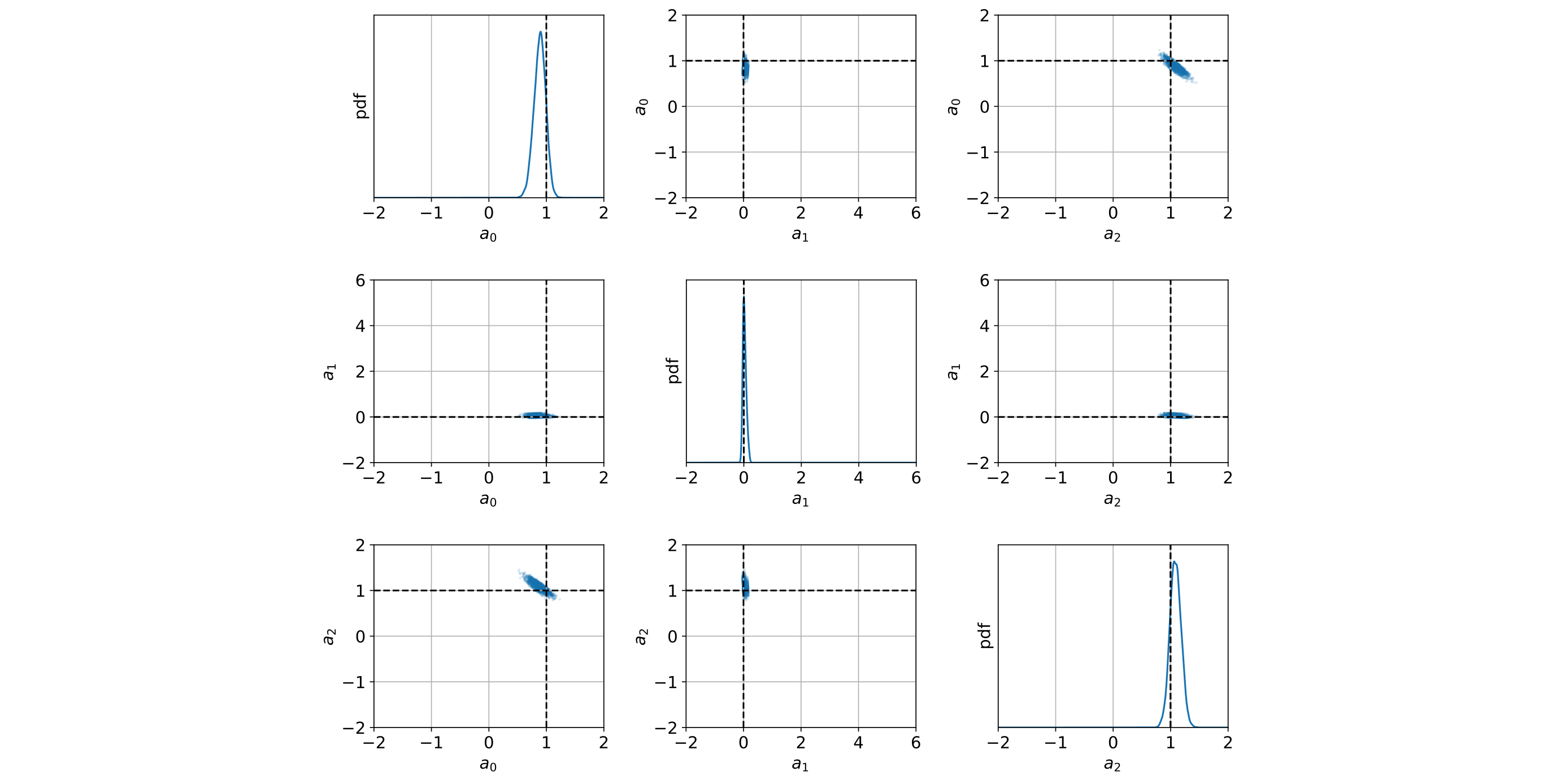}
\caption{Marginal and (pairwise) joint posterior pdfs of parameters $a_0$, $a_1$, and $a_2$ obtained using NSBL.}
\label{fig:nsbl_post_params}
\end{figure}

Recall this high precision prior on $a_1$ effectively reduces this parameter to a Dirac delta function at zero, thereby resulting in predictions of the form $y=a_0 + a_2x^2$. The removal of the uncertainty associated with parameter $a_1$ results in more precise posterior estimates of $a_0$ and $a_2$ as shown in Figure~\ref{fig:nsbl_post_params}. 

Associated with more precise posterior parameter estimates is the reduced uncertainty in the predictions of $y^*$, shown later in Figure~\ref{fig:pred_nsbl}.

\subsubsection{Hierarchical Bayesian inference}
\begin{figure}[ht!]
\centering
\includegraphics[width=\textwidth]{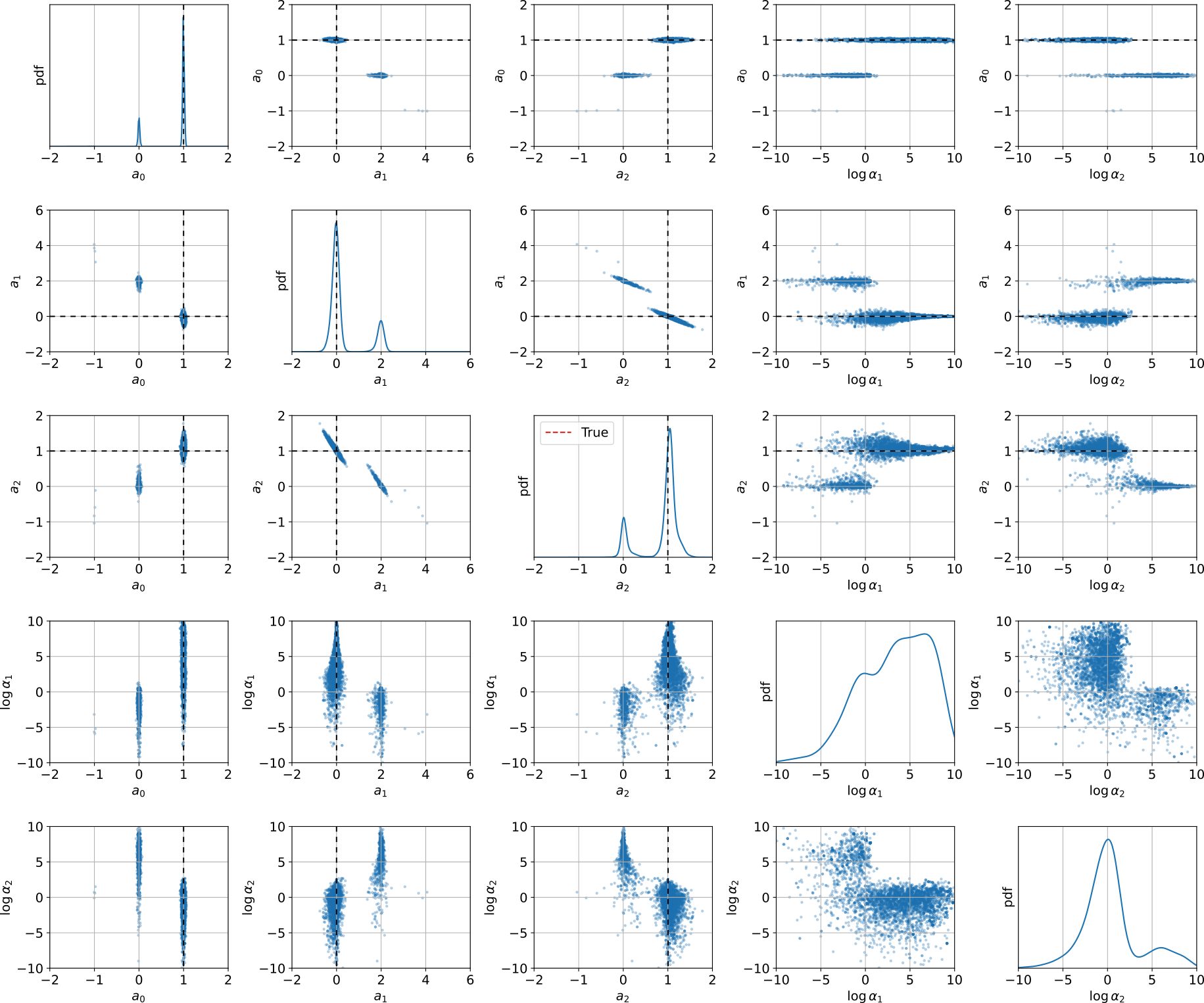}
\caption{Marginal and (pairwise) joint posterior pdfs of parameters $a_0$, $a_1$, and $a_2$ and hyperparameters $\alpha_1$, and $\alpha_2$ obtained using hierarchical Bayesian inference}
\label{fig:hier_post_params}
\end{figure}

Jointly estimating the polynomial coefficients $a_0$, $a_1$, and $a_2$ and the hyperparameters $\alpha_1$ and $\alpha_2$, whereby the questionable parameters $a_1$ and $a_2$ are assigned ARD priors, with precision $\alpha_1$ and $\alpha_2$, the inference problem can be stated as in Eq.~(\ref{2:inference}). The prior pdfs of the \textit{a priori} relevant parameters is given by Eq.~(\ref{eq:prior_a0} and the prior pdfs of the irrelevant parameters are as in Eq.~(\ref{eq:prior_phi_alpha}). The hyperprior is given by Eq.~(\ref{2:gamma}) with shape and rate parameters $r_1 = r_2 = 1+\exp(-10)$ and $s_1 = s_2 = \exp(-10)$. This parameterization of the Gamma hyperprior results in an approximately uniform distribution \textcolor{comment}{in the range} $\exp(-10) \leq \alpha_1, \alpha_2 \leq \exp(10)$ \textcolor{comment}{(see Figure \ref{fig:nsbl_hyperprior})}. Thus, the hyperparameter posterior is largely data-driven, with the hyperprior providing an upper bound on the precision of a redundant parameter. The resulting parameter posterior pdfs remain multimodal as shown in Figure \ref{fig:hier_post_params}. This can be attributed to the retention of uncertainty in the hyperparameters, as observed in Figure \ref{fig:hier_post_params}. The uncertainty in hyperparameters $\alpha_1$ and $\alpha_2$ permit various combinations of high- and low-precision priors for parameters $a_1$ and $a_2$ with non-zero probability, as evidenced by the scatter plot of samples from the joint posterior pdf in Figure \ref{fig:hier_post_params}.

\subsubsection{Comparison of \textcolor{comment}{Standard Bayesian inference,} NSBL and Hierarchical Bayesian inference }
\textcolor{comment}{In this section, we compare the predictions obtained for the Standard Bayesian inference, NSBL, and Hierarchical Bayesian inference.} The principal difference between the \textcolor{comment}{ Hierarchical Bayesian inference and NSBL} lies in their respective goals. Both methods permit the estimation of the parameters prior precision. In hierarchical Bayesian inference \textcolor{comment}{the posterior of hyperparameters are obtained}, whereas NSBL considers only the MAP estimate of \textcolor{comment}{hyperparameters}. The goal of NSBL, however, is to induce sparsity among the set of parameters $\{\bm{\phi}_\alpha\}$, balancing the trade-off between data-fit and model complexity. This represents a model selection problem. Hence, the hyperparameters function acts as a metric by which irrelevant parameters that increase model complexity without necessarily improving the data-fitting capabilities of the model can be removed.

\begin{figure}[ht!]
\begin{center}
\begin{subfigure}[b]{0.3\textwidth}
         \centering
         \includegraphics[width=\textwidth]{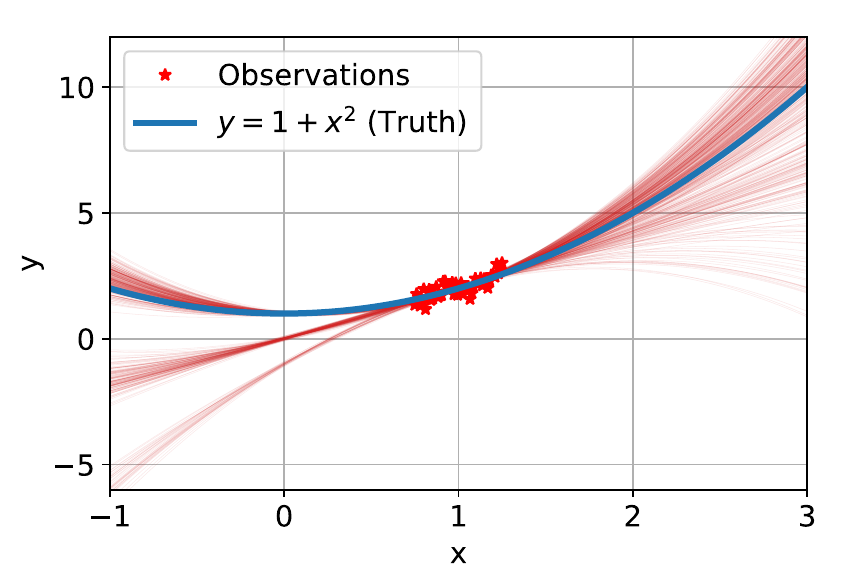}
         \caption{Standard Bayesian inference}
         \label{fig:pred_stand}
     \end{subfigure}
\begin{subfigure}[b]{0.3\textwidth}
         \centering
         \includegraphics[width=\textwidth]{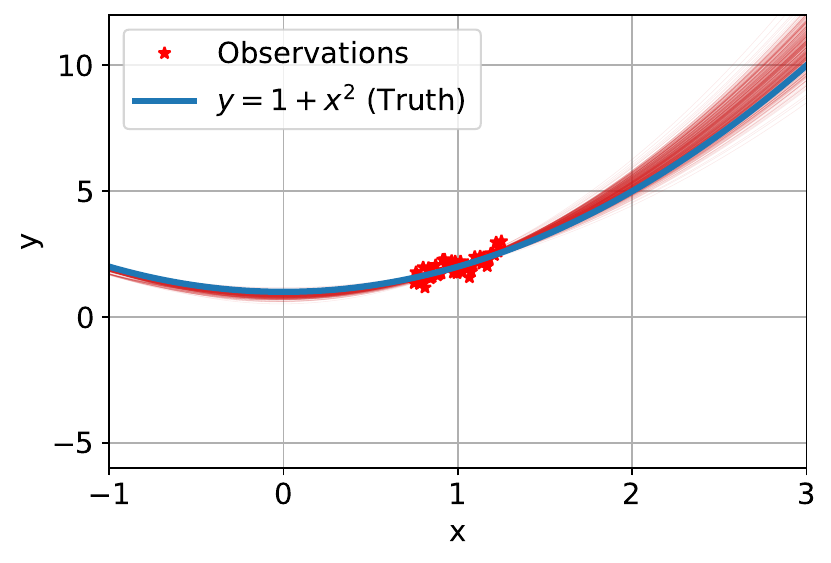}
         \caption{NSBL}
         \label{fig:pred_nsbl}
     \end{subfigure}
\begin{subfigure}[b]{0.3\textwidth}
         \centering
         \includegraphics[width=\textwidth]{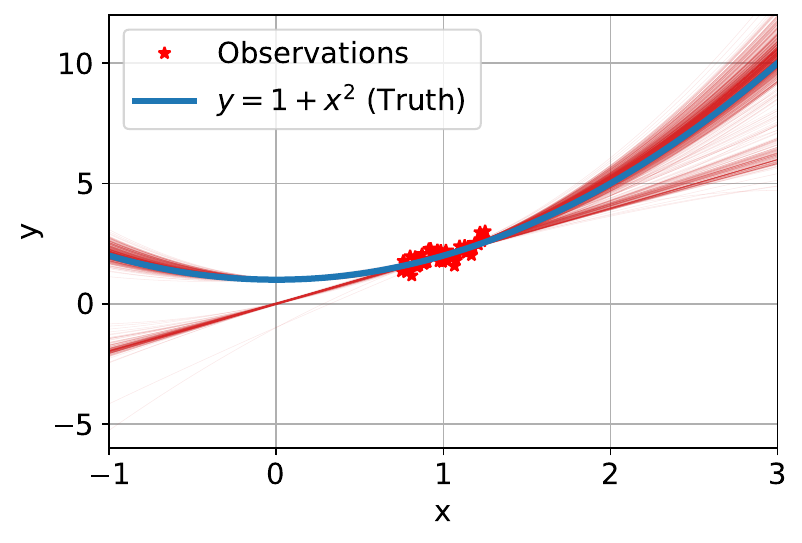}
         \caption{Hierarchical Bayesian inference}
         \label{fig:pred_hier}
     \end{subfigure}
\end{center}
\caption{Model predictions using 1000 samples from the parameter posterior pdfs. Panel (a) consists of predictions made from joint samples of $\text{p}(a_0,a_1,a_2 \vert \mathcal{D})$. Panel (b) consists  of predictions made from joint samples of $\text{p}(a_0,a_1,a_2 \vert \mathcal{D},\alpha_1^\text{map},\alpha_2^\text{map})$. Panel (c) consists of predictions made from joint samples of $\text{p}(a_0,a_1,a_2,\alpha_1,\alpha_2 \vert \mathcal{D})$}
\label{fig:predictions_summary}
\end{figure}

\begin{figure}[ht!]
\centering
\includegraphics[width=0.5\textwidth]{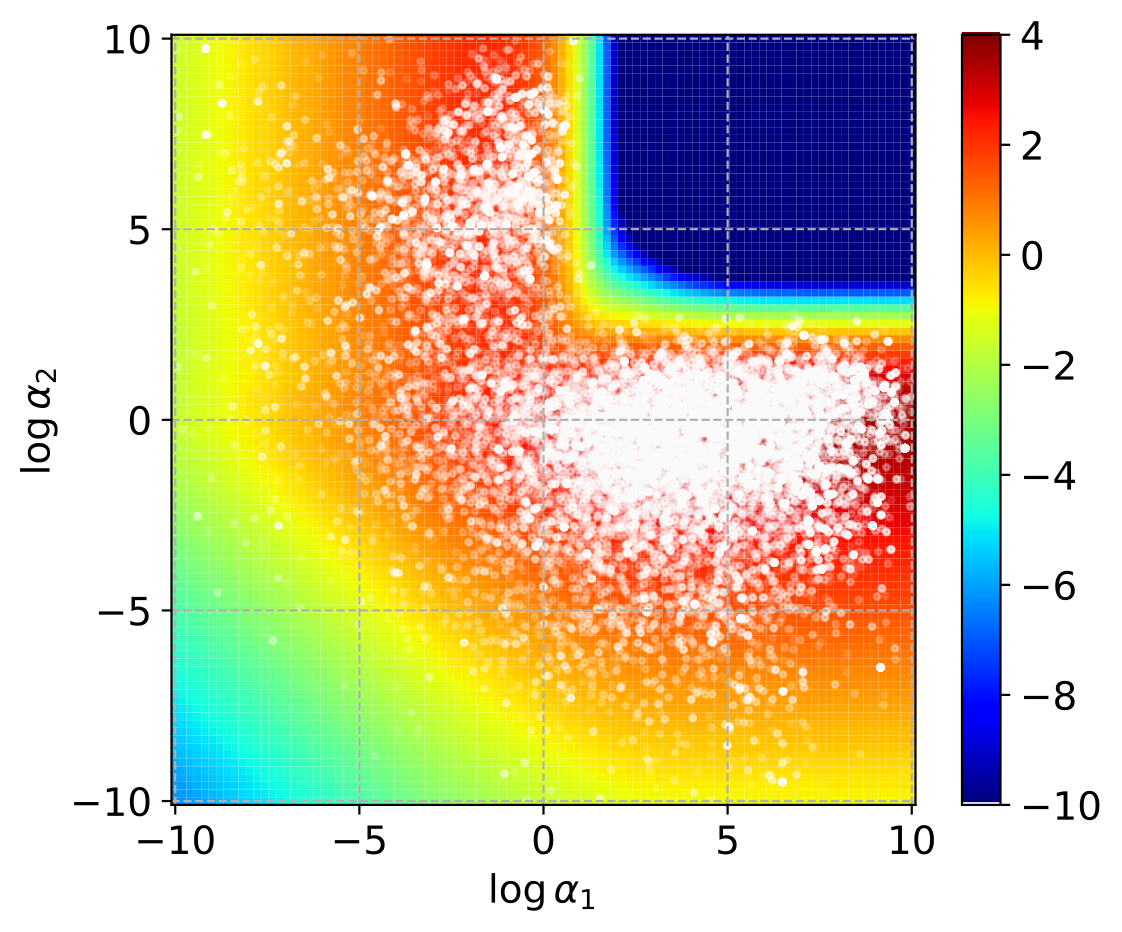}
\caption{Surface plots of the objective function for NSBL superimposed by the joint samples from the hyperparameter posterior}
\label{fig:nsbl_hierarchical}
\end{figure}

\textcolor{comment}{Figure~\ref{fig:predictions_summary} shows the prediction of standard Bayesian inference, NSBL and hierarchical Bayesian inference}. The increase in the level of hierarchy for NSBL and hierarchical Bayesian inference (see Figure~\ref{fig:pred_nsbl} and \ref{fig:pred_hier}, respectively) intuitively result in improved predictions \textcolor{comment}{in terms of uncertainty} compared to the results obtained by standard Bayesian inference \textcolor{comment}{see Figure~\ref{fig:pred_stand}}. However, the reason the NSBL, an approximate method, appears to provide \textcolor{comment}{comparable} predictions \textcolor{comment}{in terms of reduction of uncertainty} to the hierarchical Bayesian inference approach requires some attention \textcolor{comment}{being a significant aspect in this paper}. The explanation for the reduced uncertainty in the predictions can be understood by referring to the superposition of the hyperparameter posterior samples on the NSBL objective function in Figure~\ref{fig:nsbl_hierarchical}. The objective function is unimodal with a unique optimum, whereas the high-probability density region in the hyperparameter posterior exhibits an L-shape. We note that the mode of the hyperparameter posterior coincides with the global optimum of the NSBL objective function. Thus the majority of the hyperparameter posterior samples are generated from the space corresponding to large $\alpha_1$ ($a_1$ being irrelevant) and small $\alpha_2$, ($a_2$ being relevant). However, given the hierarchical Bayesian inference considers \textcolor{comment}{the entire joint posterior of the} hyperparameters, it is important to note that samples are generated from the spaces corresponding to small $\alpha_1$ and large $\alpha_2$ as well as small $\alpha_1$ and $\alpha_2$. By contrast, since NSBL considers only the MAP of the hyperparameter posterior, the forecasts made using NSBL only correspond to large $\alpha_1$ and low $\alpha_2$. The conclusions regarding the automatic sparsity inducing nature of hierarchical Bayesian inference for the linear-in-parameter model does not hold in this case. The multimodality in the hyperparameter posterior results in the the parameter posterior pdfs also remain multimodal, as evidenced through the joint samples of the parameters and hyperparameters in Figure \ref{fig:hier_post_params}. 

It is possible that the objective function becomes multimodal itself \cite{sandhu2022}, in which case there would be multiple optima. Using multistart or global optimization, it is possible to identify all modes in the objective function, however, the NSBL framework does not include any mechanism by which the knowledge of the existence of multiple optima can be reflected, as the relevance indicator depends only on the ratio of posterior to prior precision at the global optimum. Thus, while there may be multiple possible combinations of relevant and irrelevant parameters having large evidence, only the single combination which maximizes the objective function will be considered\textcolor{comment}{in NSBL}. The hierarchical Bayesian inference is more conservative as it helps improve the posterior predictions, without discarding the uncertainty in the hyperparameters.

{\color{recent}
In this example, we have examined how the inclusion of a tri-modal prior with sparse and noisy data observed over a limited range resulted in a tri-modal posterior using standard Bayesian inference. Comparing Figure \ref{fig:sbl_hierarchical} to Figure \ref{fig:nsbl_hierarchical}, the influence of the prior also has a significant effect on the shape of the hyperparameter posterior/objective function. However, as we demonstrate in Figure~\ref{fig:increase_compare}, if (a) the noise precision $\rho$ is increased, (b) the number of data points $N_d$ is increased, or (c) the range of observations is increased, the posterior estimates will reduce to a single mode. In the case where it is not possible or feasible to improve the quality of the data, we must instead focus on improving the estimation process. Increasing the level of hierarchy in our inference procedure permits the systematic removal of redundant modes from the parameter posterior pdf. This is achieved through NSBL and through a hierarchical Bayesian inference framework, whereby the non-informative prior $\text{p}(a_1,a_2)$ is replaced by an ARD prior, a zero-mean Gaussian distribution with variable precision as in Eq.~(\ref{2:logaplpha}).

\begin{figure}[ht!]
\begin{center}
\begin{subfigure}[b]{\textwidth}
         \centering
         \includegraphics[width=\textwidth]{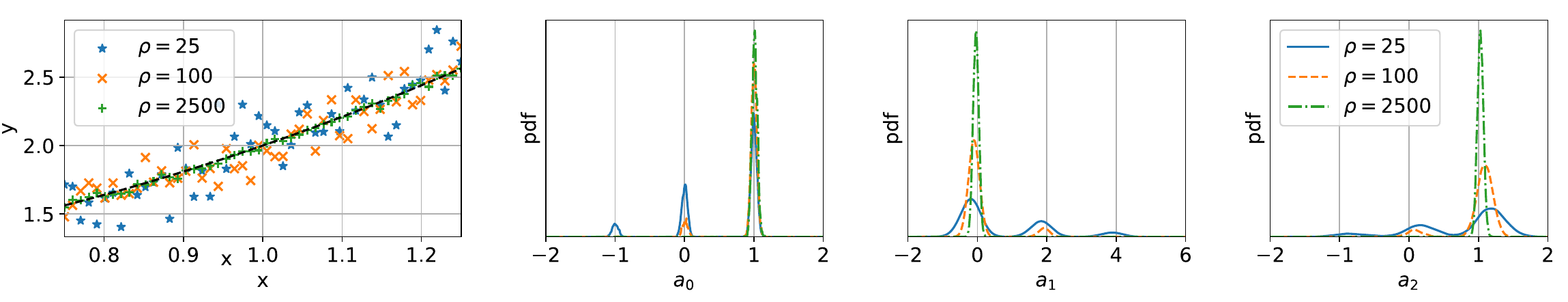}
         \caption{Increasing data noise precision, with fixed number of data points (50) and data range ($0.75 \leq x \leq 1.25$)}
     \end{subfigure}
\\
\begin{subfigure}[b]{\textwidth}
         \centering
         \includegraphics[width=\textwidth]{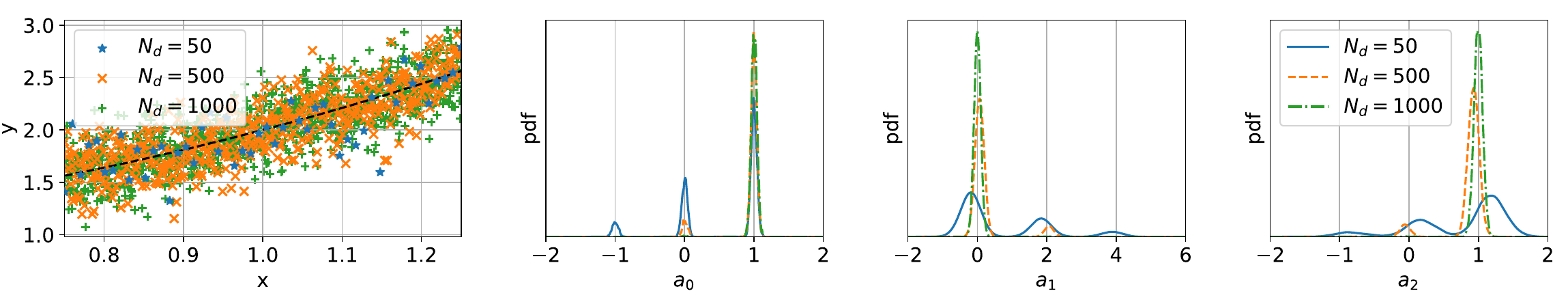}
         \caption{Increasing number of data points, with fixed noise precision (25) and data range ($0.75 \leq x \leq 1.25$)}
     \end{subfigure}
\\
\begin{subfigure}[b]{\textwidth}
         \centering
         \includegraphics[width=\textwidth]{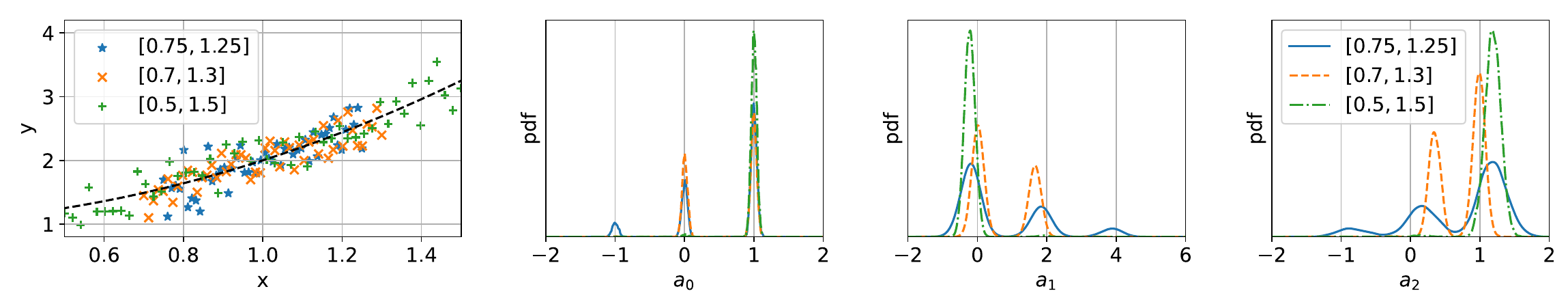}
         \caption{Increased data range, with fixed noise precision (25) and number of data points (50)}
         \label{fig:increase_range}
     \end{subfigure}
\end{center}
\caption{The number of modes in the parameter posterior distribution decreases with improved data.}
\label{fig:increase_compare}
\end{figure}

For the three cases considered in Figure \ref{fig:increase_compare}, as the quality of the data increases, the parameter posterior pdfs tend toward a single Gaussian kernel, whereby a Laplace approximation would be adequate, and MCMC sampling would no longer be required. For the case where the range of observations is increased in Figure \ref{fig:increase_range}, we demonstrate the effect of the improved data in the hyperparameter space in Figure \ref{fig:alphas_compare}. We look at both the joint hyperparameter samples from the hierarchical Bayesian setting and the objective function from NSBL. The two methodologies exhibit qualitatively similar tendencies as the range of data increases. The objective function exhibits a unique optimum and the hyperparaemter posterior tends towards a unimodal pdf (see the rightmost pannel of Figure \ref{fig:alphas_compare}).}

\begin{figure}[ht!]
\begin{center}
\begin{subfigure}[b]{\textwidth}
         \centering
          \includegraphics[width=\textwidth]{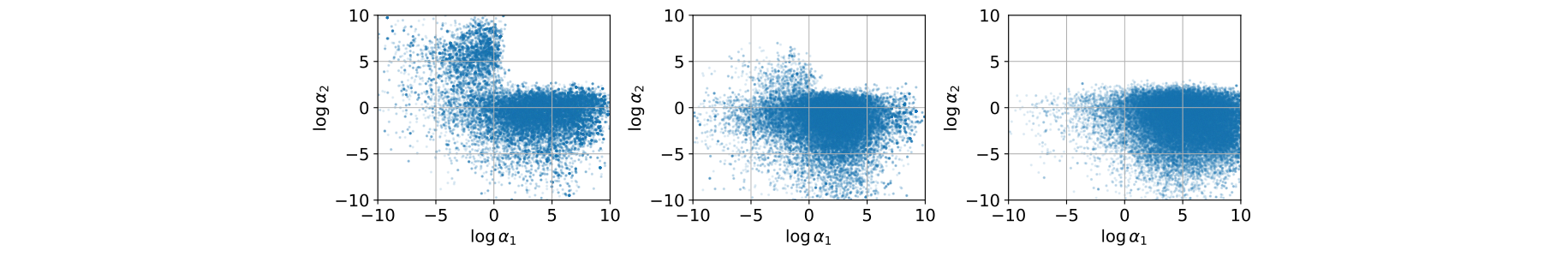}\caption{}
         \label{fig:hierarchical_alphas}
     \end{subfigure}\\
\begin{subfigure}[b]{\textwidth}
         \centering
         \includegraphics[width=\textwidth]{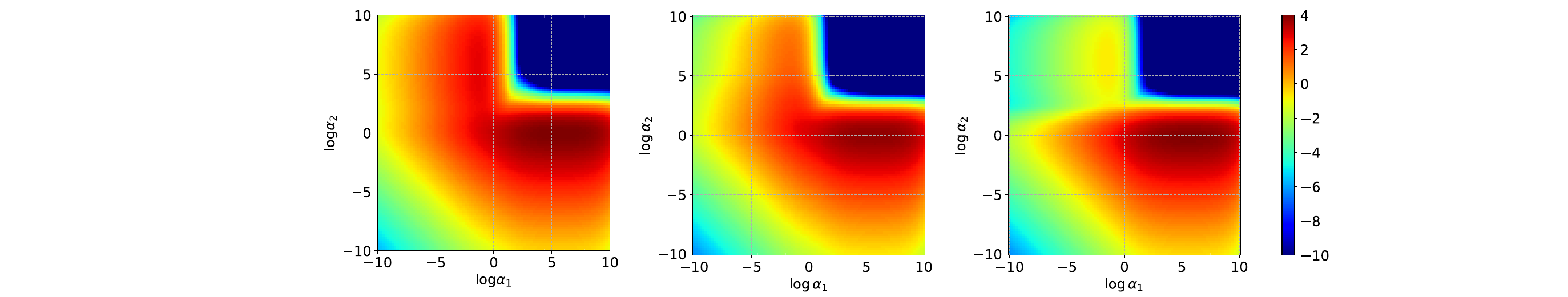}
         \caption{}
         \label{fig:nsbl_alphas}
     \end{subfigure} 
\end{center}
\caption{Increasing the range of data as shown in Figure \ref{fig:increase_range}, we comparing (a) the joint samples of the hyperparameters $\{\alpha_1,\alpha_2\}$ in the hierarchical setting against (b) the NSBL objective function in log space. The left column corresponds to data observed on [0.75,1.25], the middle column corresponds to data observed on [0.7,1.3], and the right column corresponds to data observed on [0.5,1.5].}
\label{fig:alphas_compare}
\end{figure}

\subsection{Case 3) Non-Gaussian prior and Non-Gaussian likelihood: Application of NSBL in mass-spring-damper system}
\label{sec.3.2}

\begin{figure}
\subfloat[3-dof mass-spring-damper model.]{\raisebox{-.5\height}{%
  \includegraphics[height=260pt,width=0.45\textwidth]{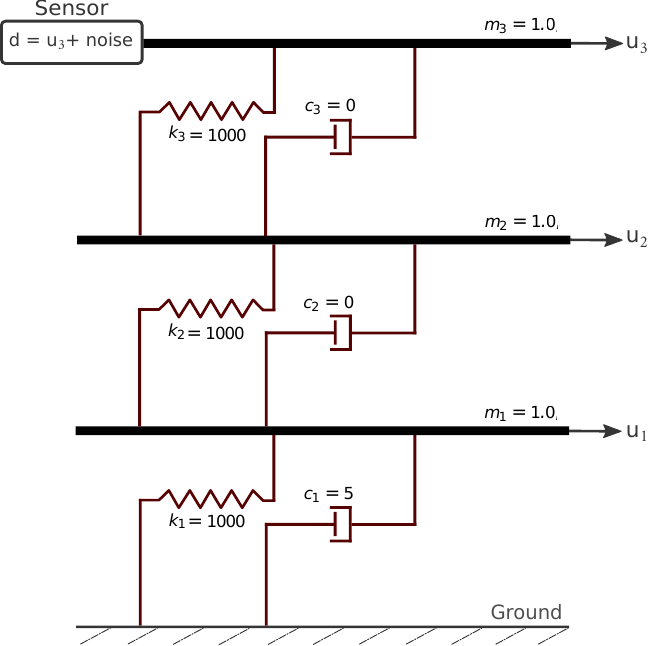}}%
  \label{fig:3dof}\rule[-25pt]{0pt}{50pt}}%
\qquad
\subfloat[Floors displacements and observation data.]{\raisebox{-.5\height}{%
  \includegraphics[height=250pt,width=0.35\textwidth]{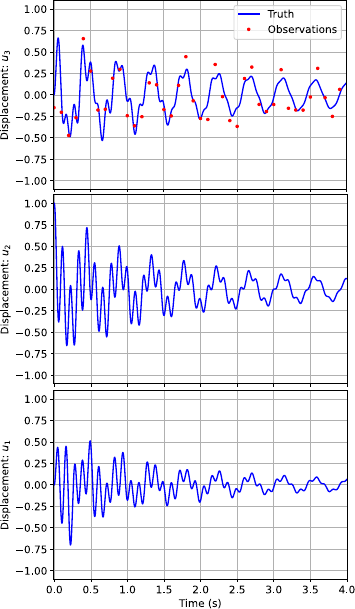}}%
  \label{fig:dips}}%
\centering
\caption{Inverse problem setup for the shear building frame}
\label{fig:3dof-output}
\end{figure}

In this particular section, we revisit a multi-storey shear building frame with rigid floors that was originally introduced by Sandhu et al. \cite{sandhu2021,sandhu2020model}. The system compromises a three-dof mass-spring-damper shown in Fig. \ref{fig:3dof}, where $\mathbf{M}$, $\mathbf{K}$, and $\mathbf{C}$ denoted as mass, stiffness, and damping matrices respectively. The equation of motion can be written as 
\begin{equation}
\mathbf{M}\ddot{\mathbf{u}} + \mathbf{C}\dot{\mathbf{u}} + \mathbf{K}\mathbf{u} = \mathbf{f}(t)
\end{equation}

\textcolor{comment}{with $\mathbf{u(t=0)} = \mathbf{u_0}$,  $\mathbf{\dot{u}(t=0)} = \mathbf{\dot{u}_0}$ }, where $\mathbf{u}, \dot{\mathbf{u}}$, and $\ddot{\mathbf{u}}$ are displacement, velocity, and acceleration vectors of the system, respectively; $\mathbf{f}(t)$ signifies external forcing, and in this specific example it is assumed to be zero. The motion then occurs by imposing an initial condition in the absence of external force in the form of oscillatory decay, as shown in Figure~\ref{fig:dips}. 

Rewriting the equation of motion in state space representation, which describes the dynamics of the building in terms of its displacements, we have
\begin{equation}
\begin{aligned}\label{statespace}
    \dot{\mathbf{x}}(t) = \mathbf{A}\mathbf{x}(t)     
\end{aligned}
\end{equation}
where $\mathbf{x} = \{\mathbf{u},
\dot{\mathbf{u}}\}$ is the state vector and
$\mathbf{A} =\mathbf{[0, I; -M^{-1}K, -M^{-1}C]}$ noted as the system matrix describes how the state variables are changing over time. The responses of the structural system can then be obtained by solving the first-order ordinary differential equation given in Eq.~\ref{statespace}. The following values are defined to simulate the three-dof system:  the initial condition is set as $\mathbf{x_0} = \{0,1,0,0,0,0\}$,  mass for each dof is $m_1 = m_2 = m_3 = 1.0$, the stiffness values are $k_1 = k_2 = k_3 = 1000.0$, and damping coefficients are specified as $c_1 = 5$, $c_2 = 0$, and $c_3 = 0$. Given the initial state $\mathbf{x(t)}=\mathbf{x_0}$, the solution is derived as $\mathbf{x}(t) = e^{At}\mathbf{x_0} $.

In order to pose an inverse problem, a noisy observation is generated by introducing an additive Gaussian noise $\mathbf{\epsilon} \sim \mathcal{N}(0,0.1)$ to the actual response of the third-floor displacement ($u_3$). This includes 40 data points collected over a duration of four seconds and represented as dots in Fig. \ref{fig:dips} (top panel). Now, we attempt to estimate each of the inter-storey damping and stiffness coefficients, $\bm{\phi}=\left\{c_1, c_2, c_3, k_1, k_2, k_3\right\}$. 

 No prior knowledge of damping parameters is assumed. Therefore, they are treated as \textcolor{comment}{ questionable parameters denoted as $\bm{\phi}_{\alpha} = \left\{ c_1, c_2, c_3\right\}$ to which we assign a non-informative prior.} On the other hand, \textcolor{comment}{the stiffness coefficients are considered as \textit{a priori} relevant parameters denoted as $\bm{\phi}_{-\alpha}=\left\{ k_1, k_2, k_3\right\}$  which are strictly positive with the prior pdf given by}

\begin{equation}\label{eq:3dof_known}
\mathrm{p}\left(\bm{\phi}_{-\alpha}\right) = \mathcal{U}\left(k_1 \mid 0,2000\right)\mathcal{U}\left(k_2 \mid 0,2000\right) \mathcal{U}\left(k_3 \mid 0,2000\right),
\end{equation}

\begin{figure}[ht!]
\centering
\includegraphics[width=\textwidth]{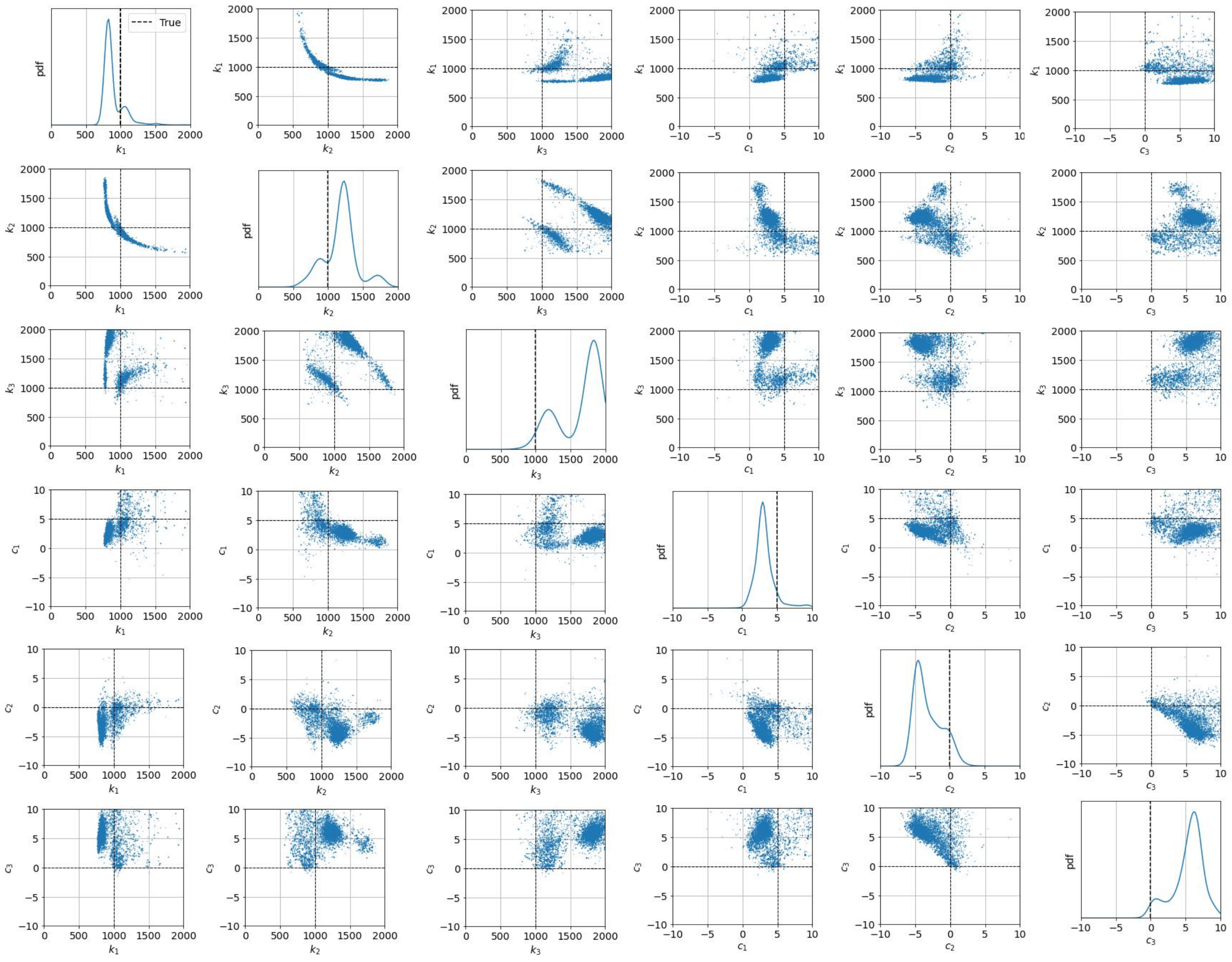}
\caption{Marginal and (pairwise) joint posterior pdfs of stiffness and damping parameters  obtained using standard Bayesian inference}
\label{fig:3dof-marginal-3dof}
\end{figure}

In the standard Bayesian setup, the joint posterior pdfs of stiffness coefficients $\left\{ k_1, k_2, k_3\right\}$ and damping parameters $\left\{ c_1, c_2, c_3\right\}$ are obtained using TMCMC to generate samples from the partial posterior pdf ${\mathrm{p}(\mathcal{D} \mid \phi) \mathrm{p}\left(\phi_{-\alpha}\right)}$ for the GMM as per Eq. \ref{2:GMM}. The marginal pdfs pertaining to this KDE approximation and pairwise joint posterior pdfs of these parameters are shown in Figure~\ref{fig:3dof-marginal-3dof}. \textcolor{acc}{Notice that despite the use of a linear structural dynamics model, the partial posterior pdfs are non-Gaussian, which is caused by the nonlinear relation between the unknown parameters (damping and stiffness coefficients) and the observations (displacement at the third storey).} Additionally, the presence of sparse and noisy data results in the emergence of multimodality in the posterior pdfs, which \textcolor{comment}{differs from the results provided in Sandhu et al. \cite{sandhu2021}.} The non-Gaussian features in the posterior distributions are mainly caused by the multimodality in the likelihood function, which is, in turn, caused by the noisy and sparse data. 

Furthermore, there is a significant high probability region located at a distance from the true values, as shown by dashed lines in Figure~\ref{fig:3dof-marginal-3dof}. Most notably, the available sparse and noisy data do not provide enough information to estimate the damping coefficients accurately, leading to a negative value for $c_2$ (being physically unrealistic). While the current data is relatively small, sparse, and noisy, this leads to the lack of robustness in the model. Additionally, the presence of sparse, noisy, and incomplete data in conjunction with an overparameterized model contributes to overfitting, as will be shown later (Figure~\ref{fig:3dof-prediction-disp} and \ref{fig:3dof-prediction-vel}). \textcolor{acc}{This overfitting of over-parameterized models can be improved by data enhancement or by reducing the level of noise in the data. However, in real-life scenarios, this may not be possible.}  Therefore, we instead focus on improving the estimation process. \textcolor{comment}{Increasing the level of hierarchy in our inference procedure permits the systematic removal of redundant modes from the parameter posterior pdfs. This is achieved through the hierarchical Bayesian Inference and NSBL, where the non-informative prior for damping parameters $(c_1, c_2, c_3)$ are replaced by an ARD prior and will be discussed in the following sections}. 


\subsubsection{Nonlinear Sparse Bayesian Learning}
Following the idea of hybrid prior discussed in Section~\ref{sec2.2.1}, $\bm{\phi}$ is decomposed into the questionable parameter vector $\bm{\phi}_\alpha=\left\{c_1, c_2, c_3\right\}$, and \textit{a priori} relevant parameter vector $\bm{\phi}_{-\alpha}=\left\{k_1, k_2, k_3\right\}$. The ARD prior assigned to the damping coefficients $\bm{\phi}_\alpha=\left\{c_1, c_2, c_3\right\}$, which is defined as,

\begin{equation}\label{eq:3dof_questionable}
\mathrm{p}\left(\boldsymbol{\phi}_\alpha \mid \boldsymbol{\alpha}\right) = \mathcal{N}\left(c_1 \mid 0, \alpha_1^{-1}\right) \mathcal{N}\left(c_2 \mid 0, \alpha_2^{-1}\right) \mathcal{N}\left(c_3 \mid 0, \alpha_3^{-1}\right).
\end{equation}

Combining Eqs.~(\ref{eq:3dof_known}) and (\ref{eq:3dof_questionable}) gives the hybrid prior for the system. The hyperparameters priors $\mathrm{p}(\boldsymbol{\alpha})$ is defined as Eq.~\ref{2:logaplpha} with $\log r_i = \log s_i = -10$. 

\begin{figure}[ht!]
\begin{subfigure}[b]{\textwidth}
\begin{center}
\begin{subfigure}[b]{0.3\textwidth}
         \centering
         \includegraphics[width=\textwidth]{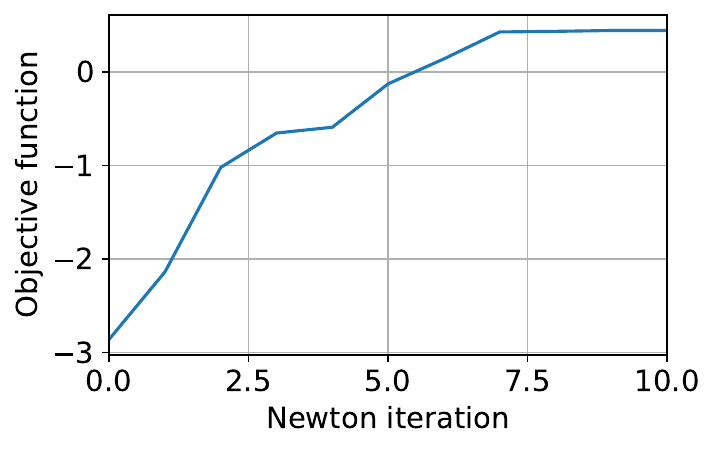}
     \end{subfigure}
\begin{subfigure}[b]{0.3\textwidth}
         \centering
         \includegraphics[width=\textwidth]{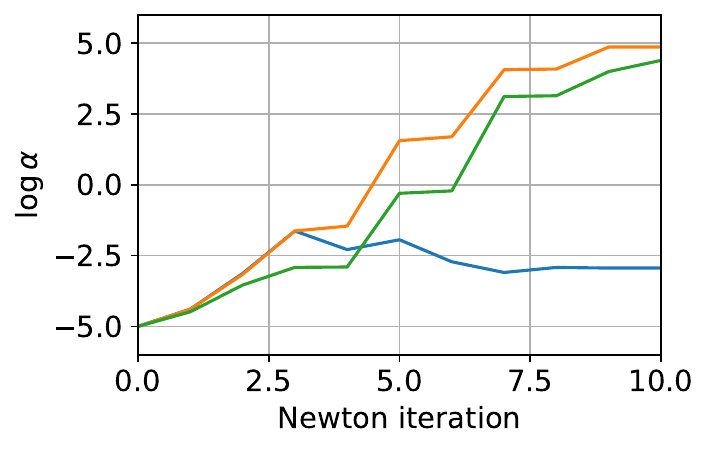}
     \end{subfigure}
\begin{subfigure}[b]{0.3\textwidth}
         \centering
         \includegraphics[width=\textwidth]{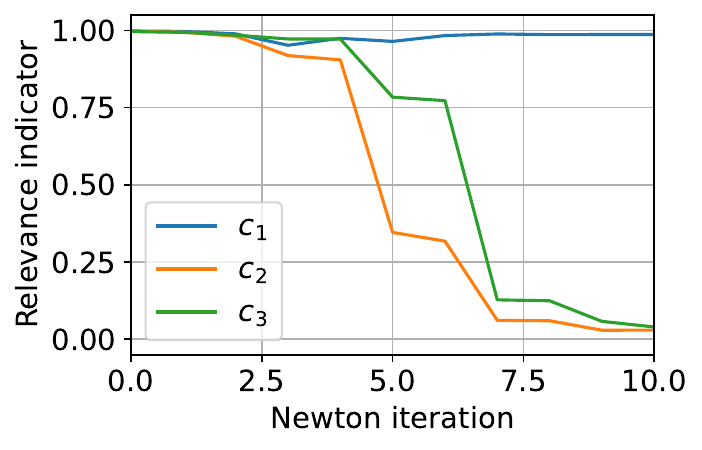}
     \end{subfigure}
\end{center}
\caption{$\boldsymbol{\alpha}$ initiated at \{-5,-5,-5\}}
\label{fig:3dof_optimization_1}
\end{subfigure}
\\
\begin{subfigure}[b]{\textwidth}
\begin{center}
\begin{subfigure}[b]{0.3\textwidth}
         \centering
         \includegraphics[width=\textwidth]{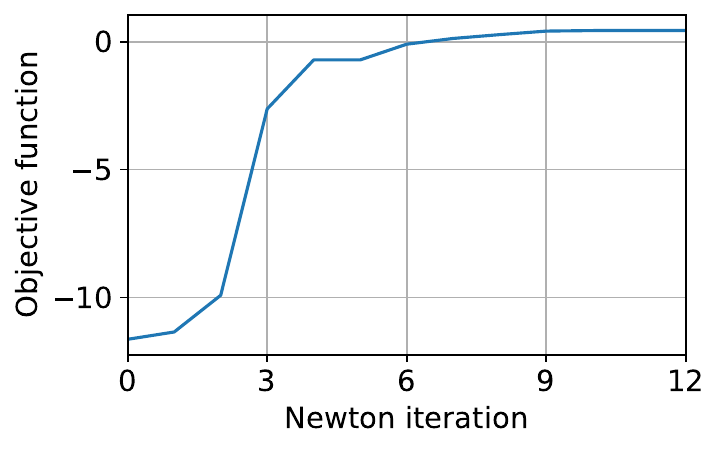}
     \end{subfigure}
\begin{subfigure}[b]{0.3\textwidth}
         \centering
         \includegraphics[width=\textwidth]{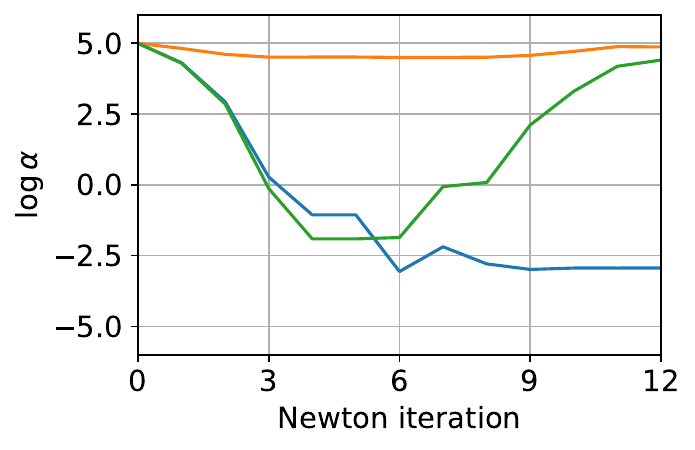}   
     \end{subfigure}
\begin{subfigure}[b]{0.3\textwidth}
         \centering
         \includegraphics[width=\textwidth]{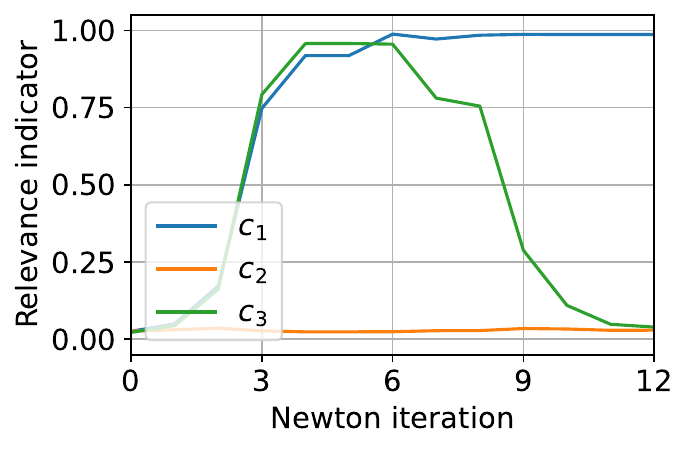}
     \end{subfigure}
\end{center}
\caption{$\boldsymbol{\alpha}$ initiated at initiated at \{5,5,5\}}
\label{fig:3dof_optimization_2}
\end{subfigure}
\\
\begin{subfigure}[b]{\textwidth}
\begin{center}
\begin{subfigure}[b]{0.3\textwidth}
         \centering
         \includegraphics[width=\textwidth]{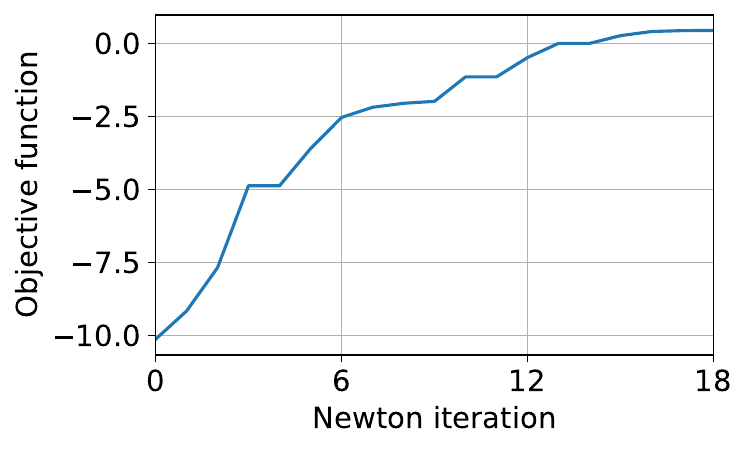}
     \end{subfigure}
\begin{subfigure}[b]{0.3\textwidth}
         \centering
         \includegraphics[width=\textwidth]{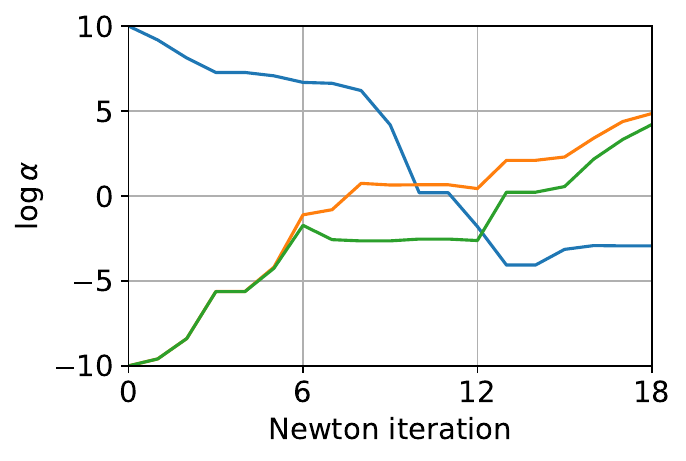}   
     \end{subfigure}
\begin{subfigure}[b]{0.3\textwidth}
         \centering
         \includegraphics[width=\textwidth]{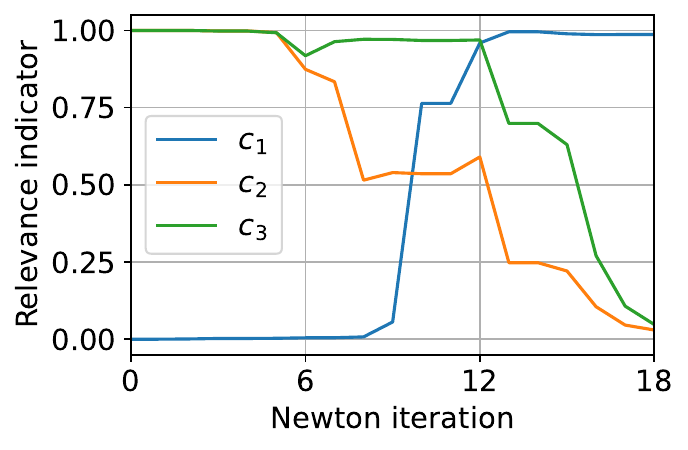}
     \end{subfigure}
\end{center}
\caption{$\boldsymbol{\alpha}$ initiated at \{10,-10,-10\}}
\label{fig:3dof_optimization_3}
\end{subfigure}
\caption{NSBL optimization as a function of Newton's iterations with different start points}
\label{fig:3dof_optimization}
\end{figure}

In order to calculate the sparse representation of damping coefficients, a multistart Newton iteration is initiated following the NSBL algorithm discussed in Section \ref{sec.2.2}. Figure~\ref{fig:3dof_optimization} demonstrates the convergence of the optimization algorithm as a function of Newton's iteration for three different choices of starting $\log{\boldsymbol{\alpha}}$ values. Through the semi-analytical NSBL framework, we obtain the optimal value of $\log \alpha_1$, $\log \alpha_2$, and $\log \alpha_3$ equal to \{-2.93, 4.87, 4.40\}, respectively. As the hyperparameters $\boldsymbol{\alpha}$ appear in the precision of prior, the large values for $\alpha_2$, and $\alpha_3$ correspond to the damping parameters $c_2$ and $c_3$, effectively constrain these parameters converge towards a Dirac delta function centered at zero. In summary, \textcolor{comment}{for various initialization of $\boldsymbol{\alpha}$,} the relevance indicator reveals the relevance of the first-floor damping $c_1$ and irrelevance of the second and third-floor damping $c_2$ and $c_3$ matching the data-generating model. 

\begin{figure}[ht!]
\begin{subfigure}[b]{\textwidth}
\begin{center}
\begin{subfigure}[b]{0.28\textwidth}
         \centering
         \includegraphics[width=\textwidth]{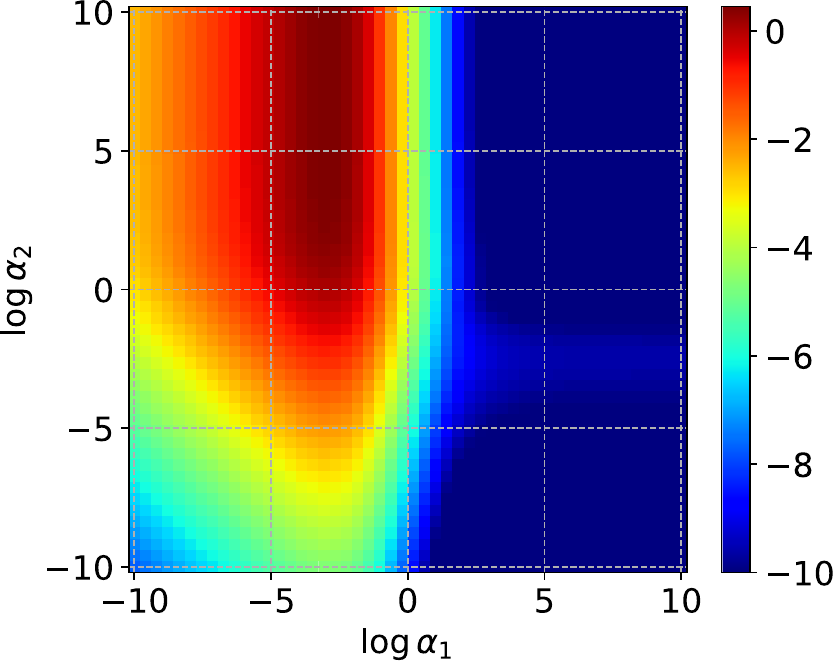}
     \end{subfigure}
\begin{subfigure}[b]{0.28\textwidth}
         \centering
         \includegraphics[width=\textwidth]{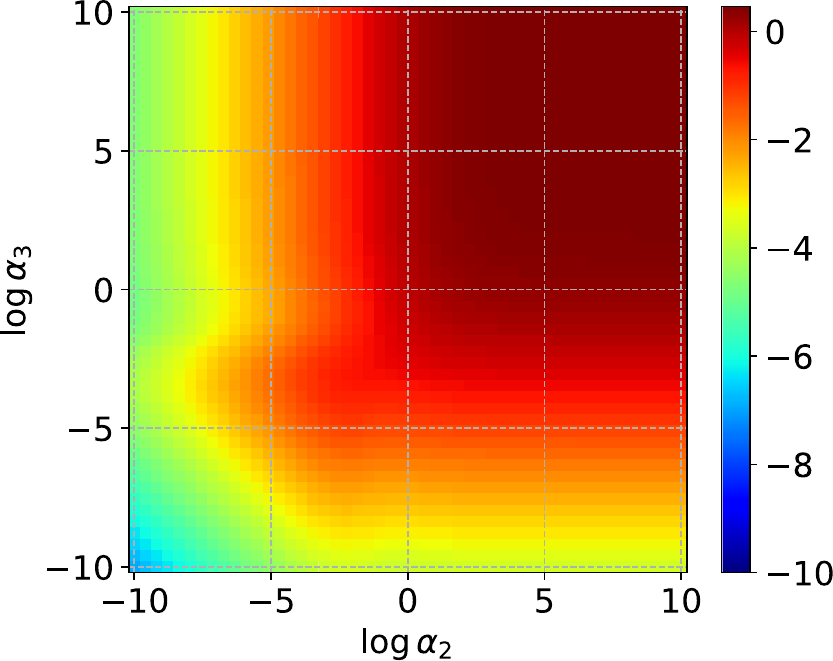}
     \end{subfigure}
\begin{subfigure}[b]{0.28\textwidth}
         \centering
         \includegraphics[width=\textwidth]{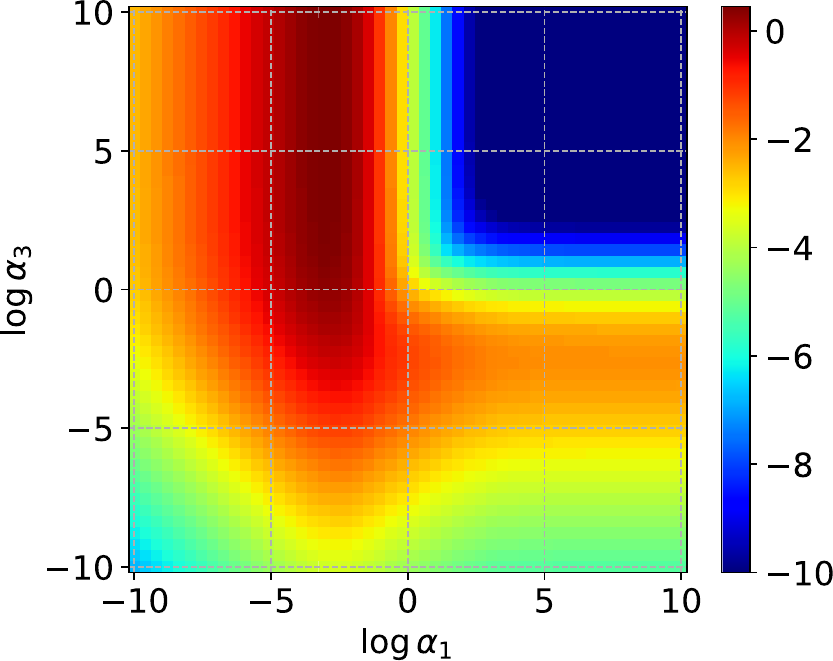}
     \end{subfigure}
\end{center}
\caption{Model evidence }
\label{3dof_surf_model evidence}
\end{subfigure}
\\
\begin{subfigure}[b]{\textwidth}
\begin{center}
\begin{subfigure}[b]{0.28\textwidth}
         \centering
         \includegraphics[width=\textwidth]{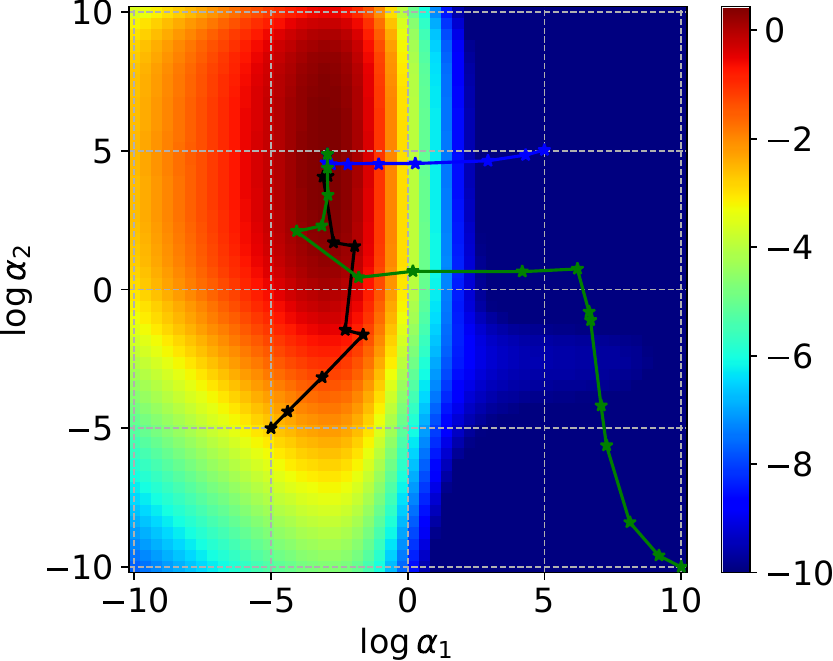}
     \end{subfigure}
\begin{subfigure}[b]{0.28\textwidth}
         \centering
         \includegraphics[width=\textwidth]{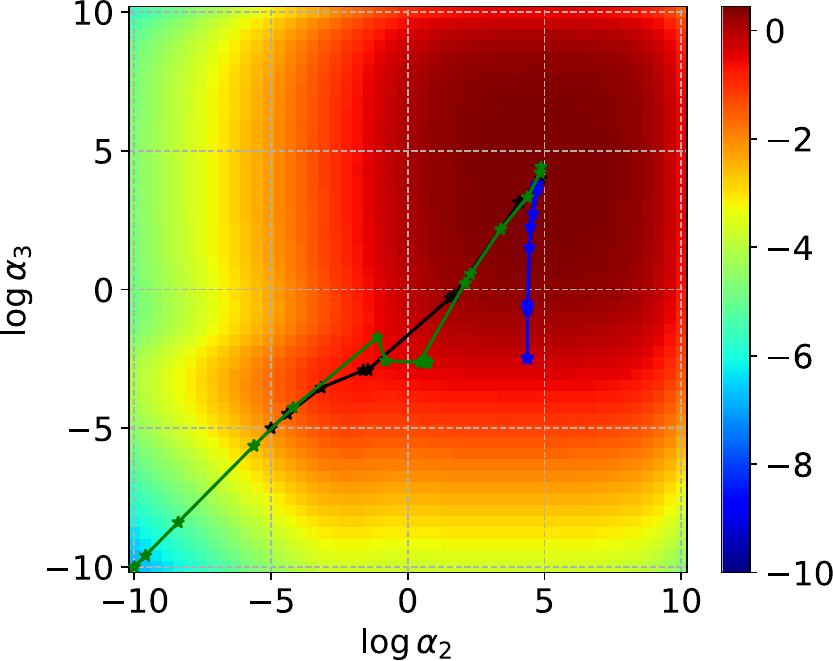}   
     \end{subfigure}
\begin{subfigure}[b]{0.28\textwidth}
         \centering
         \includegraphics[width=\textwidth]{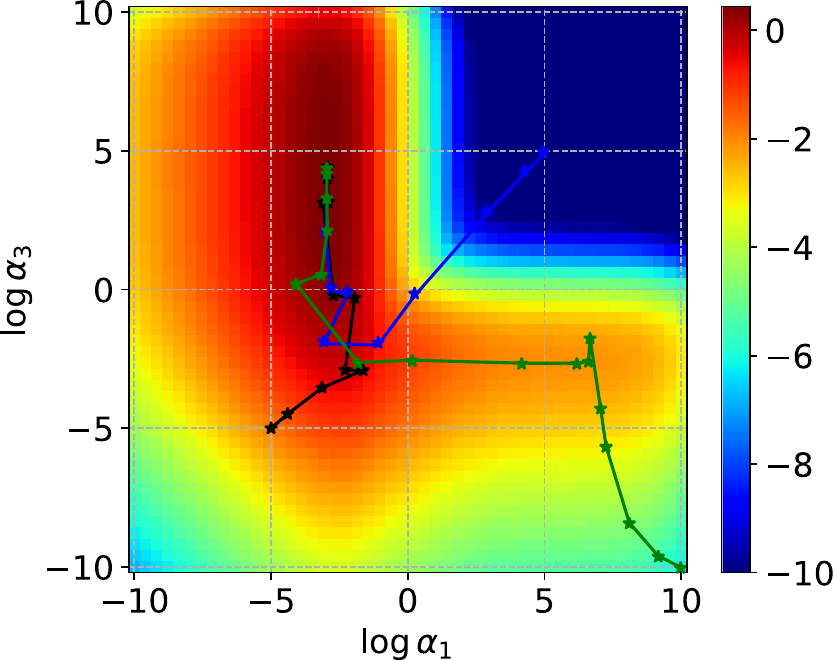}
     \end{subfigure}
\end{center}
\caption{Objective function }
\label{3dof_surf_objective function}
\end{subfigure}
\caption{Surface plots of the model evidence and objective function for NSBL as a function of $\log(\mathbf{\alpha})$  superimposed by Newton's iterations with different start points for $\boldsymbol{\alpha}$; black-initiated at \{-5,-5,-5\}, blue-initiated at \{5,5,5\} shown in blue, green-initiated at \{10,-10,-10\}}
\label{fig:3dof-nsbl_surfs}
\end{figure}
In order to gain some insight into the behavior of the NSBL's objective function, Figure \ref{fig:3dof-nsbl_surfs}, \textcolor{comment}{depicts the estimate of} the model evidence and the resulting objective function as a function $\alpha$ parameter. \textcolor{comment}{It is important to note that, in this particular problem, the NSBL cost function is a function of three precision parameters $\alpha_1,\alpha_2$, and $\alpha_3$ and visualizing its optimal points can be challenging. Hence, we generate a two-dimensional plot of its corresponding model evidence and objective function varying two of ${\alpha}$ parameters while fixing the third ${\alpha}$ parameter at its corresponding MAP estimate. Note that its global optimum provides the MAP estimate of $\boldsymbol{\alpha}$, denoted by $\boldsymbol{\alpha}^\text{MAP} = \left\{\alpha_1^\text{MAP},\alpha_2^\text{MAP},\alpha_3^\text{MAP}\right\}$. While the global optimum of these two-dimensional plots of the objective functions does not coincide with the MAP estimate, they still provide valuable information on the characteristic of objective function affecting Newton's iteration in the optimization process.} Newton's iteration with various utilization of $\boldsymbol{\alpha}$ start points are also superimposed on these objective function. While there is no clear optimum in the log evidence plots in Figure~\ref{3dof_surf_model evidence}, the objective function in Figure~\ref{3dof_surf_objective function} is unimodal with a unique optimum. This results in a unique solution of $\log \boldsymbol{\alpha}$ for three different choices of starting values which is equal $\log\boldsymbol{\alpha} = \{-2.93, 4.87, 4.40\}$ obtained from multistart Newton's iterations. 

\begin{figure}[ht!]
\centering
\includegraphics[width=\textwidth]{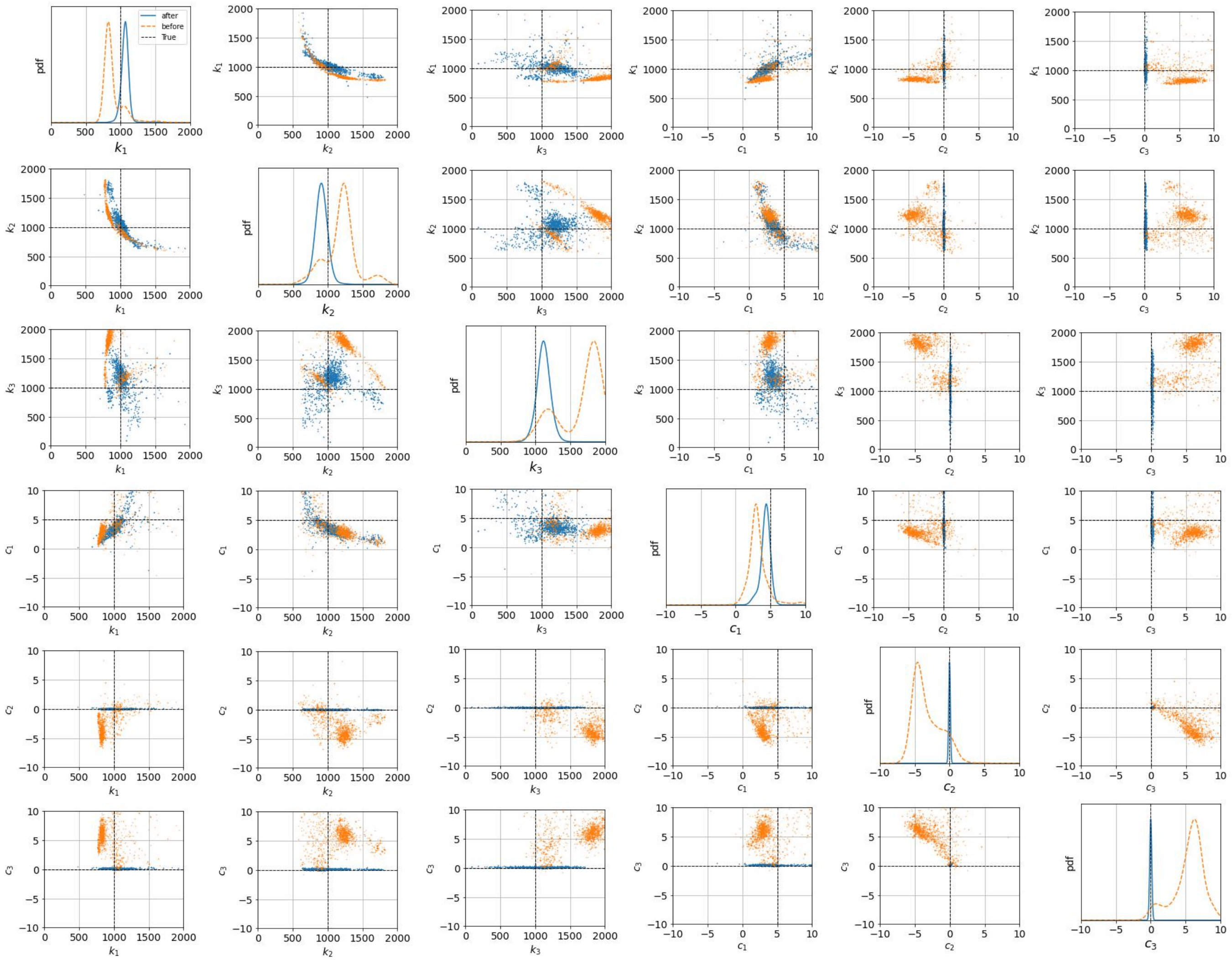}
\caption{Marginal and (pairwise) joint posterior pdfs of stiffness and damping parameters obtained using NSBL. The label \textit{before} (dashed curve) indicates the results obtained by standard Bayesian inference with non-informative priors; this is equivalent to the NSBL results before incorporating the  effect of hyperparameters. The label \textit{after} (solid curve) indicates the results after optimizing the hyperparameters using NSBL.}
\label{fig:NSBL_post_params}
\end{figure}

Finally, the marginal posterior pdfs of stiffness coefficients and damping parameters and pairwise joint posterior pdfs of these parameters, before and after sparse learning through the NSBL algorithm, are shown in Figure~\ref{fig:NSBL_post_params}. Clearly, the redundant damping parameters $c_2$ and $c_3$ are highly peaked at zero and therefore pruned off. Moreover, the modes of the marginal pdfs approach the true parameter values for other parameters.

\subsubsection{Hierarchical Bayesian inference}
\begin{figure}[ht!]
\centering
\includegraphics[width=\textwidth]{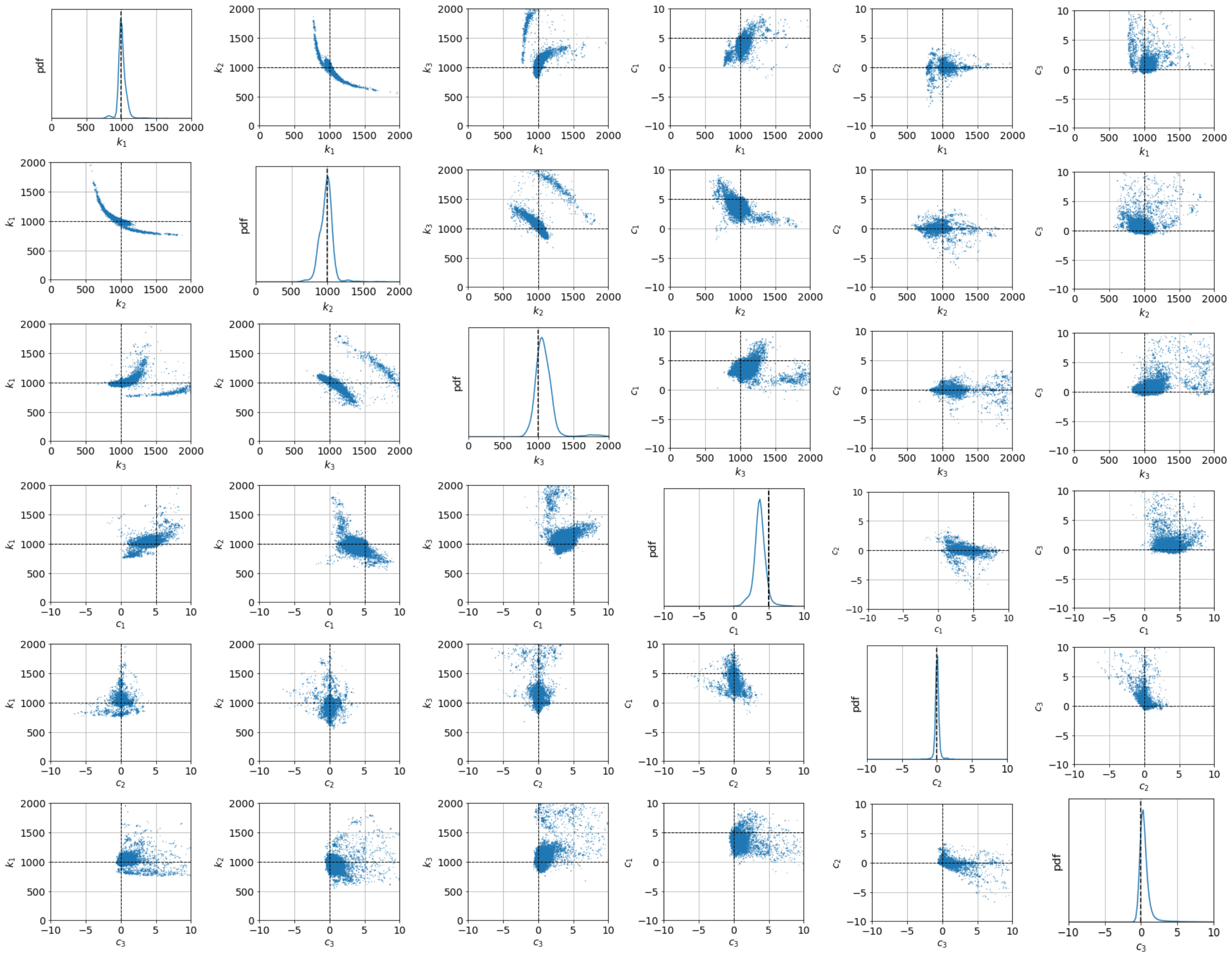}
\caption{Marginal and (pairwise) joint posterior pdfs of stiffness and damping parameters obtained using hierarchical Bayesian inference}
\label{fig:3dof_mpdf_hier}
\end{figure}

\begin{figure}[ht!]
\centering
\includegraphics[width=0.5\textwidth]{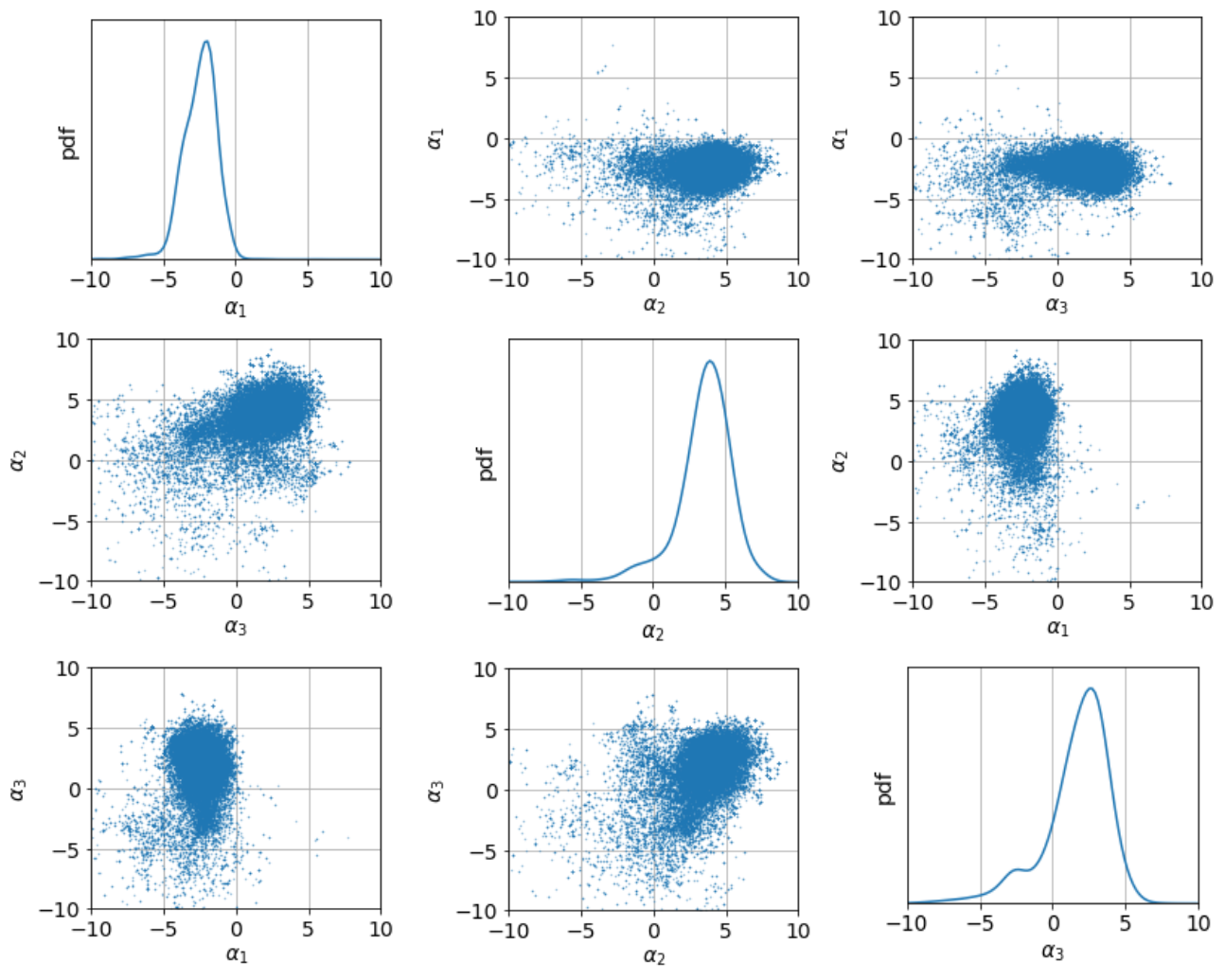}
\caption{Marginal and (pairwise) joint posterior pdfs of hyperparameter $\alpha_1$, $\alpha_2$ and $\alpha_3$ obtained using hierarchical Bayesian inference}
\label{fig:3dof_mpdf_hier_hyp}
\end{figure}

\begin{figure}[ht!]
\centering
\includegraphics[width=\textwidth]{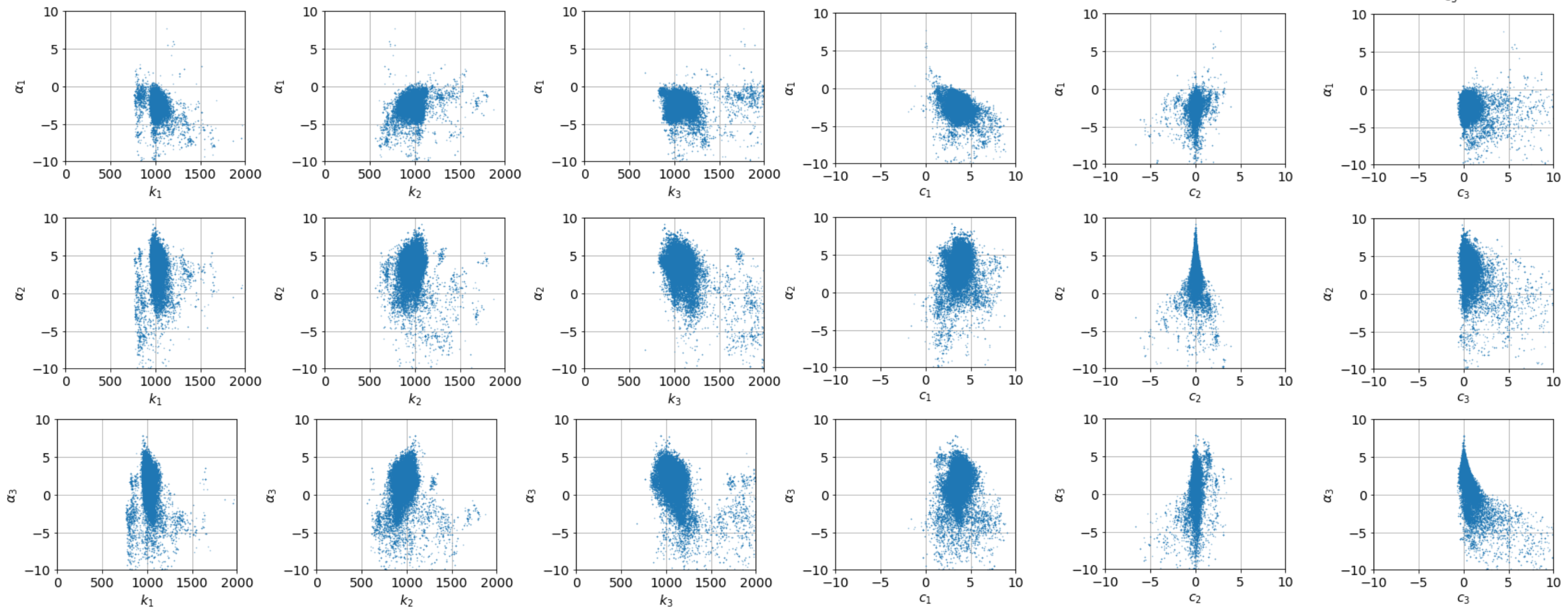}
\caption{Pairwise joint posterior pdfs of hyperparameter $\alpha_1$, $\alpha_2$ and $\alpha_3$ obtained using hierarchical Bayesian inference}
\label{fig:3dof_mpdf_hier_alpha}
\end{figure}

While NSBL depends on the MAP estimation of $\boldsymbol{\alpha}$ parameter, for hierarchical Bayesian inference, the inference problem is defined as the joint estimation of stiffness, damping coefficients, and hyperparameters \textcolor{comment}{$\left\{k_1,k_2,k_3,c_1,c_2,c_3, \alpha_1, \alpha_2, \alpha_3\right\}$}. \textcolor{comment}{The hyperprior is given by a Gamma distribution introduced in Eq. \ref{2:gamma} for $\alpha_1, \alpha_2$ and $\alpha_3$ with the shape and rate parameters $\log s_i = \log r_i = -10$ (same used for the NSBL).} \textcolor{acc}{The resulting parameter posterior pdfs along with the pairwise joint posterior pdfs for the damping coefficients ($c_1, c_2, c_3$) and the stiffness coefficients ($k_1, k_2, k_3$) are shown in Figure~\ref{fig:3dof_mpdf_hier}, displaying multimodality in parameters. The non-Gaussian posterior pdfs of the hyperparameters ($\alpha_1, \alpha_2, \alpha_3$) are plotted in Figure \ref{fig:3dof_mpdf_hier_hyp}. Moreover, the pairwise joint posterior pdfs of hyperparameters~($\alpha_1, \alpha_2, \alpha_3$) with stiffness and damping coefficients shown in Figure \ref{fig:3dof_mpdf_hier_alpha} also exhibit non-Gaussian features.}

\begin{figure}[ht!]
\begin{center}
\begin{subfigure}[b]{0.25\textwidth}
         \centering
         \includegraphics[width=\textwidth]{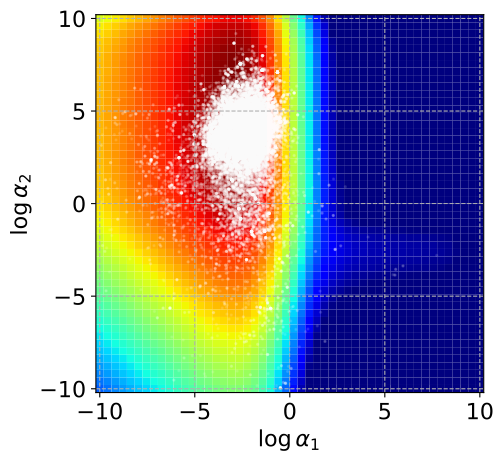}
     \end{subfigure}
\begin{subfigure}[b]{0.25\textwidth}
         \centering
         \includegraphics[width=\textwidth]{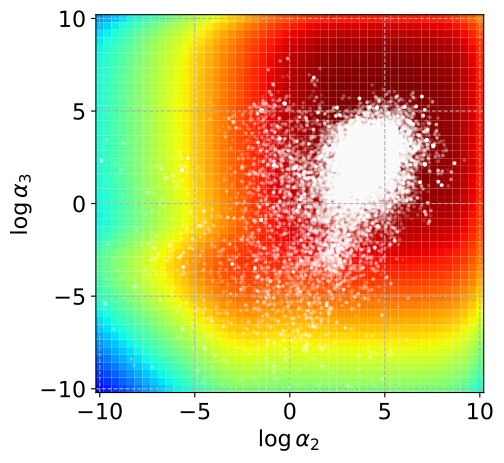}
     \end{subfigure}
\begin{subfigure}[b]{0.25\textwidth}
         \centering
         \includegraphics[width=\textwidth]{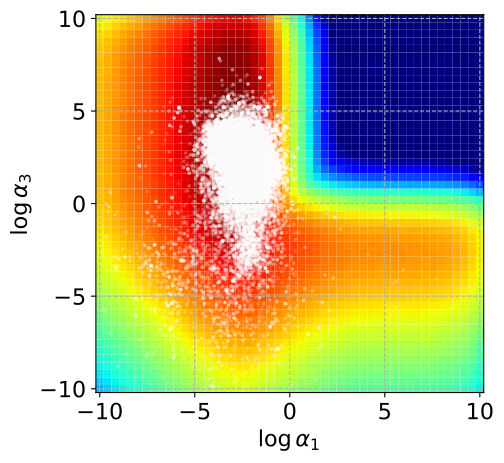}
\end{subfigure}
\caption{Surface plots of the objective function for NSBL superimposed by the joint samples from the hyperparameter posterior}
\label{fig:3dof_pdf_superimpose}
\end{center}
\end{figure}



\subsubsection{Prediction comparison of Standard Bayesian inference, NSBL, and Hierarchical Bayesian inference}
\begin{figure}[ht!]
\begin{center}
\begin{subfigure}[b]{0.3\textwidth}
         \centering
         \includegraphics[width=\textwidth]{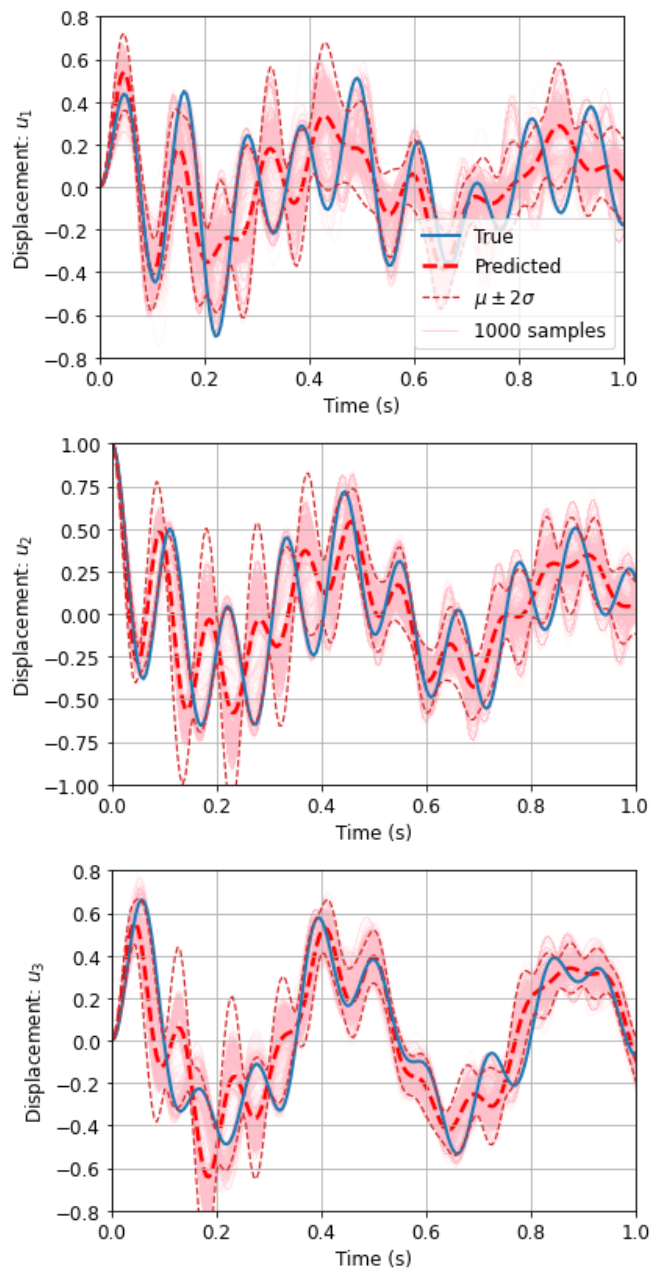}
         \caption{Standard Bayesian inference}
         \label{fig:pred-disp-bayes}
     \end{subfigure}
\begin{subfigure}[b]{0.3\textwidth}
         \centering
         \includegraphics[width=\textwidth]{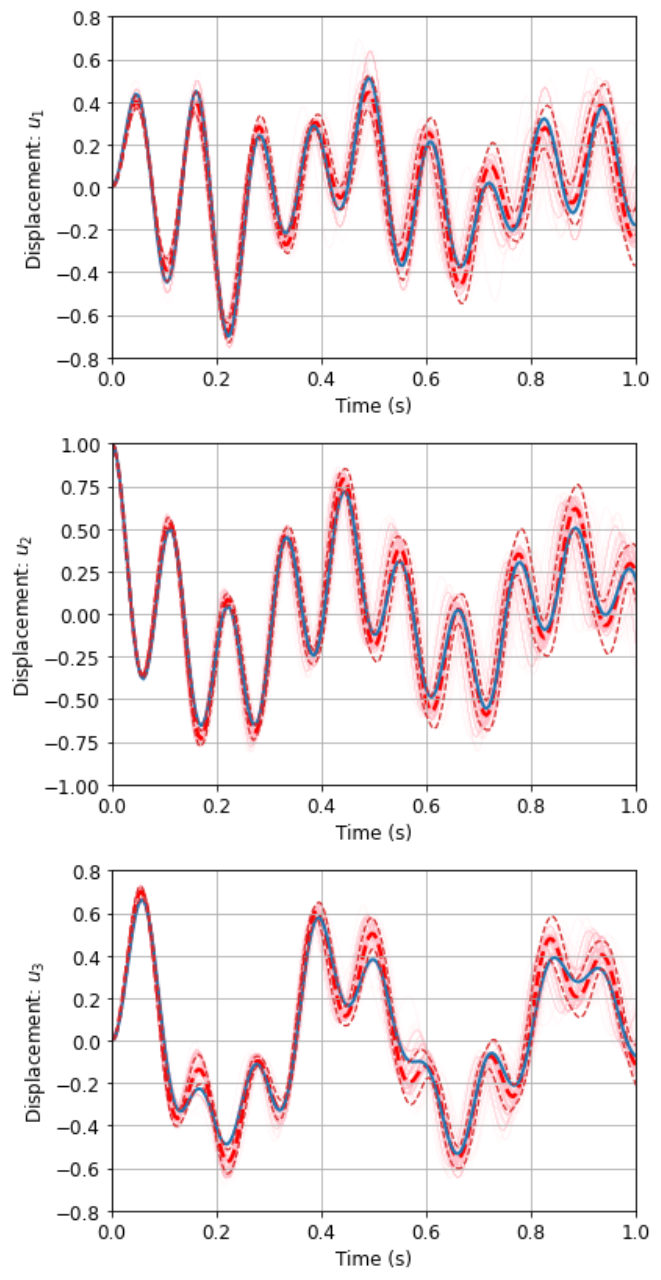}
         \caption{NSBL}
         \label{fig:pred-disp-NSBL}
     \end{subfigure}
\begin{subfigure}[b]{0.3\textwidth}
         \centering
         \includegraphics[width=\textwidth]{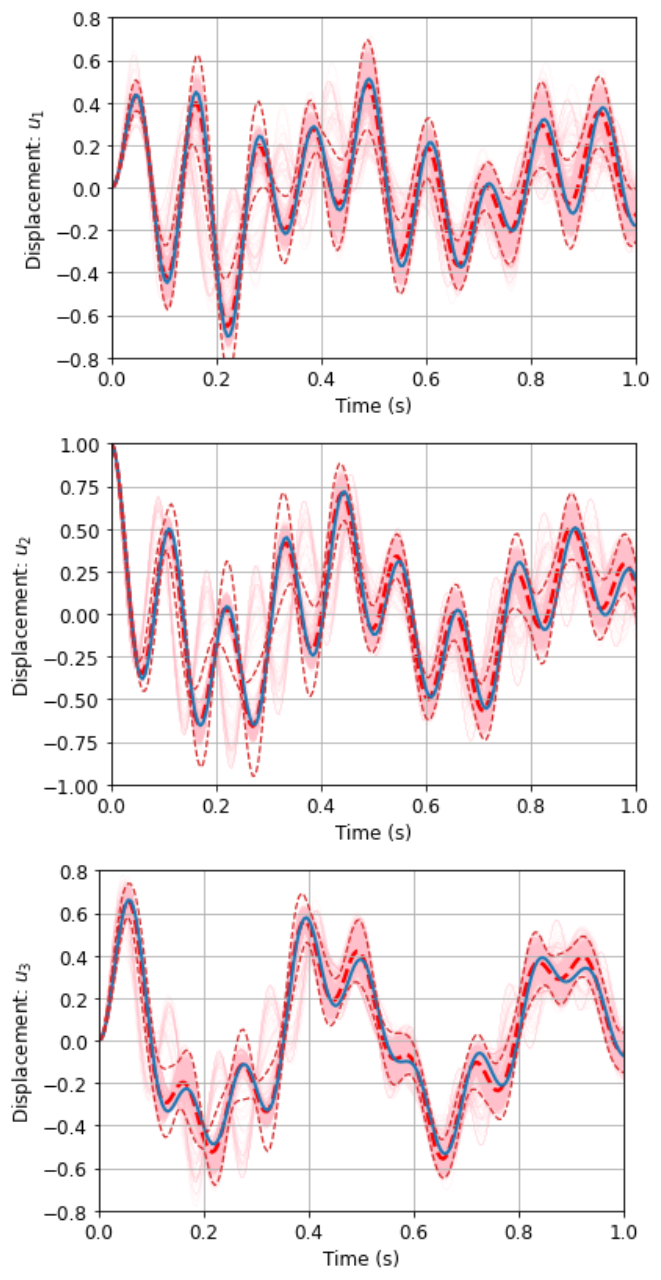}
         \caption{Hierarchical Bayesian inference}
         \label{fig:pred-disp-hier}
     \end{subfigure}
\end{center}
\caption{Predicted versus true time-history of displacement at all storeys}
\label{fig:3dof-prediction-disp}
\end{figure}

\begin{figure}[ht!]
\begin{center}

\begin{subfigure}[b]{0.3\textwidth}
         \centering
         \includegraphics[width=\textwidth]{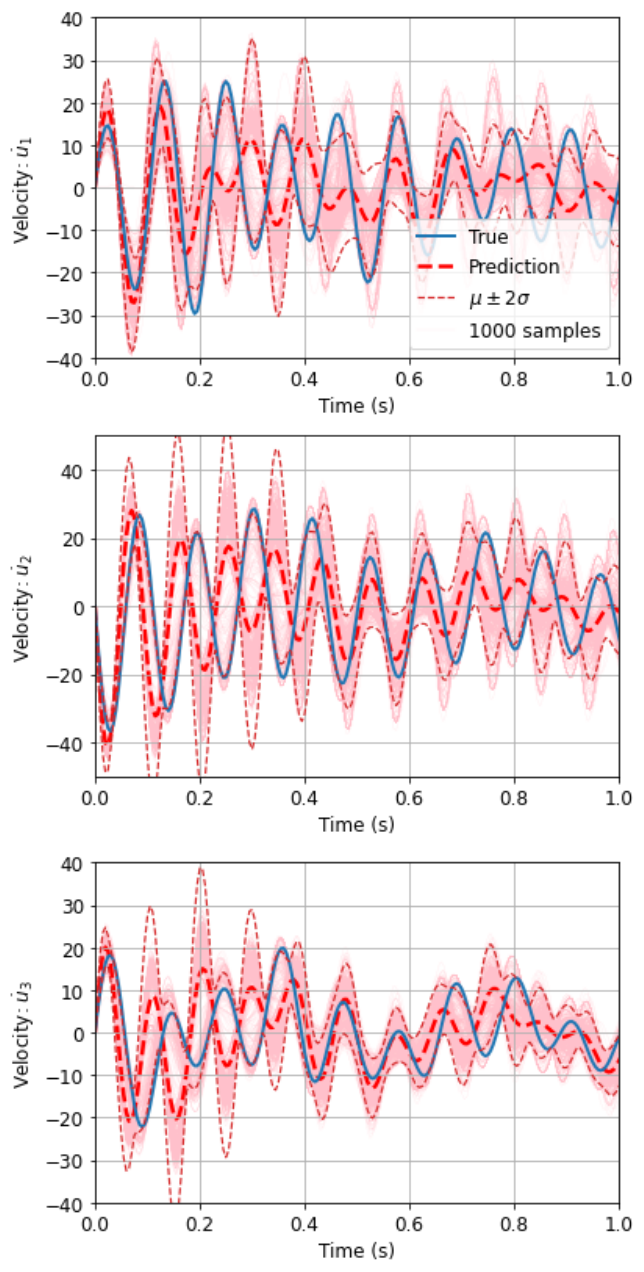}
         \caption{Standard Bayes}
         \label{fig:pred-vel-bayes}
     \end{subfigure}
\begin{subfigure}[b]{0.3\textwidth}
         \centering
         \includegraphics[width=\textwidth]{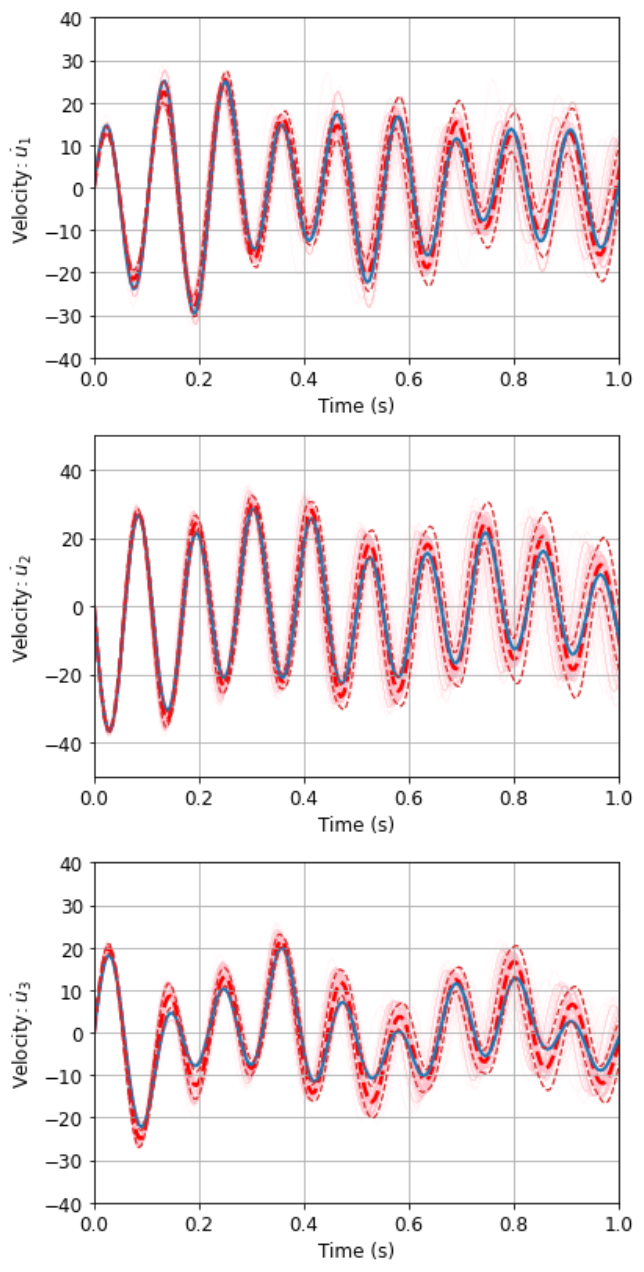}
         \caption{NSBL}
         \label{fig:pred-vel-NSBL}
     \end{subfigure}
\begin{subfigure}[b]{0.3\textwidth}
         \centering
         \includegraphics[width=\textwidth]{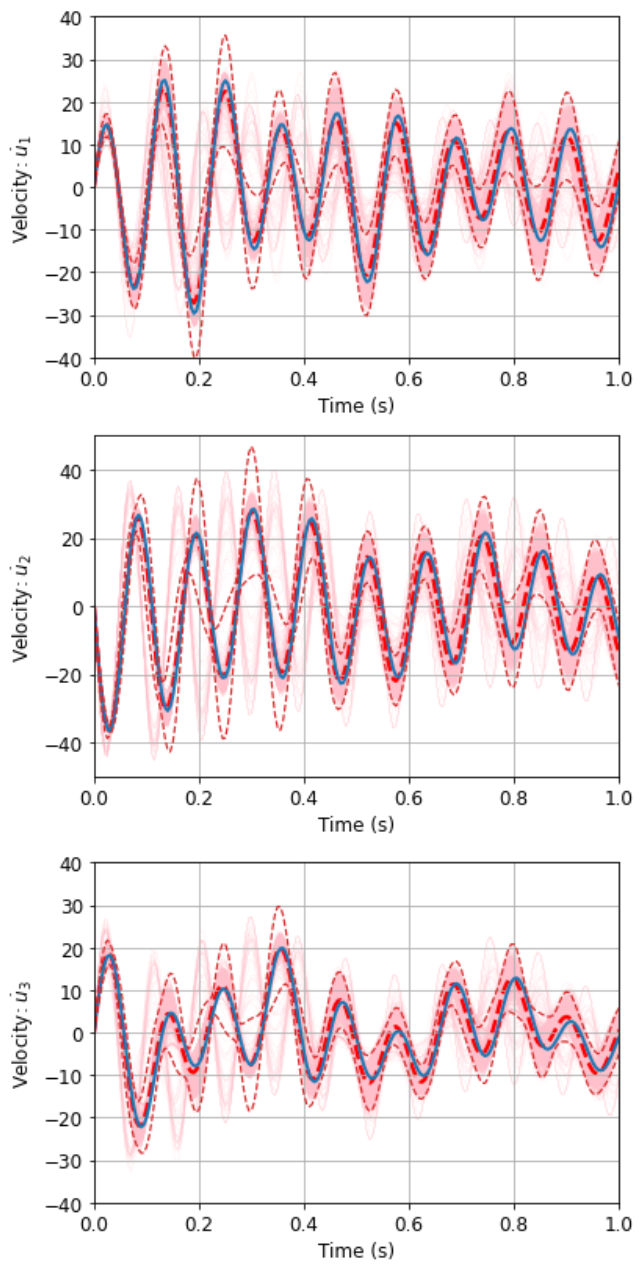}
         \caption{Hierarchical Bayes}
         \label{fig:pred-vel-hier}
     \end{subfigure}
\end{center}
\caption{Predicted versus true time-history of velocity at all storeys}
\label{fig:3dof-prediction-vel}
\end{figure}

The standard Bayesian approach provides a practical framework for modeling time series data and making predictions under uncertainty. However, with the presence of redundant parameters in the overparameterized model (in conjugation with sparse and noisy data), this approach is incapable of capturing the complexity of the underlying process and shows high uncertainty in the prediction as depicted in Figure~\ref{fig:pred-disp-bayes} for displacement and Figure~\ref{fig:pred-vel-bayes} for velocity, respectively. In contrast, NSBL and hierarchical Bayesian inference reduce uncertainty in prediction, as evident in Figures~\ref{fig:pred-disp-NSBL} and \ref{fig:pred-disp-hier}, and Figures~\ref{fig:pred-vel-NSBL} and \ref{fig:pred-vel-hier} for displacement and velocity, respectively. \textcolor{acc}{While the results of NSBL and hierarchical Bayesian inference are comparable, NSBL incurs a significantly lower computational cost (approximately half as much in terms of time for this particular example) when compared to the hierarchical Bayesian inference}.
To evaluate the accuracy of the prediction of each model, Table \ref{MEA} presents a comparison of these results for each time series with mean absolute error (MAE) ($MAE=\frac{1}{n} \sum_{i=1}^n\left|y_i-\hat{y}_i\right|$ where $y_i$ represents the actual value and $\hat {y}_i$ represents the predicted value for data point i). The lower the MAE, the better the predictive capability of the model. Note that, NSBL outperforms standard Bayesian inference and produces a comparable result to that of hierarchical Bayesian inference. 

\begin{table}[ht!]
  \caption{MAE of time-series forecasting results for each floor obtained using standard Bayesian inference, NSBL, and hierarchical Bayesian inference}
  \centering 
    \begin{tabular}{cccc}\label{MEA}
    Methods  & Standard Bayesian inference & NSBL & Hierarchical Bayesian inference\\
    \small{Metric} & MAE & MAE & MAE\\
    \hline
    displacement $u_1$  &   0.0817 &  {0.0404} & 0.0333 \\
    
     displacement $u_2$  & 0.1031 & {0.0560} & 0.0477 \\ 
    
     displacement $u_3$  & 0.0631 & {0.0393} & 0.0313 \\ 
    
    velocity $\dot{u}_1$  &  4.013 & {2.178} & 1.802 \\ 
    
     velocity $\dot{u}_2$  & 5.433 & {3.076} & 2.523 \\ 
    
     velocity $\dot{u}_3$  & 2.839 & {1.992} & 1.276 \\    
    \end{tabular}
  \end{table}

\section{Conclusion}
\label{sec:conclusion}

\textcolor{acc}{The construction of the GMM approximation of the product of the likelihood function and the informative prior serves as the cornerstone to enable a semi-analytical framework in the NSBL setting. This framework facilitates the iterative computation of Bayesian entities as functions of the hyperparameters, including the model evidence, the parameter posterior, and the objective function (and thus the gradient vector and Hessian matrix). Through some specific numerical investigations, it has been demonstrated that this methodology emerges as a competitive choice for modeling scenarios characterized by non-Gaussian priors and non-Gaussian likelihoods or both, effectively demonstrating the applicability of the NSBL framework beyond simple problems initially demonstrated in the NSBL paper by Sandhu et al. \cite{sandhu2021}. The accuracy and efficacy of this semi-analytical framework has been validated through numerical investigations, drawing comparison with hierarchical Bayesian inference. More specifically, within this context, hierarchical Bayesian inference serves dual purposes. It not only provides a reference against which the NSBL results can be compared, it also offers a deeper insight into the validity of the approximation inherent to the NSBL algorithm. This is achieved by examining the joint posterior distribution of both parameters and hyperparameters obtained through hierarchical Bayesian inference. This is a subtle, yet important point about validating the GMM-based objective function in NSBL against the hyperparameter posterior obtained by MCMC sampling in hierarchical Bayesian inference.} 

Note that, in this paper, we investigated a set of scenarios where the pdf of the likelihood times the known prior was well-sampled, thus leading to a good quality GMM approximation using KDE. We have yet to test the performance of NSBL in scenarios where the GMM is a poor representation of the partial posterior, which may arise due to sampling errors or due to an inadequate number of kernels in the case of multimodal or other highly non-Gaussian pdfs. It remains to be seen whether the integration over the parameters when computing the evidence permits effective sparse learning in the presence of a poor GMM approximation.


\textcolor{acc}{As mentioned before, NSBL is inherently approximate, as it seeks the type-II maximum a posteriori (MAP) estimate of the hyperparameters rather than the full joint posterior of the hyperparameters. This is in contrast to hierarchical Bayesian inference which provides a full account of the uncertainty in the model parameters and hyperparameters. Through numerical examples, we have illustrated the effect of eliminating the uncertainty in the hyperparameters in terms of the predictive distribution of the model outpus as well as the parameter posterior distributions. Ultimately, these results illustrated that given a sufficient number of samples are used, similar sparsity levels are obtained for both NSBL and hierarchical Bayesian inference. Moreover, the numerical investigations demonstrated the accuracy of NSBL for inverse modelling involving multimodality in both parameters and hyperparameters.} 

\textcolor{acc}{A noteworthy practical benefit of NSBL compared to hierarchical Bayesian inference is eliminating the need for sampling in the hyperparameter space. In doing so, NSBL retains the advantage of standard Bayesian inference, wherein the cost of sampling is lower than for hierarchical Bayesian inference. Furthermore, the NSBL algorithm retains sparsity-inducing capabilities of increasing the level of the hierarchy as it obtains point estimates for the hyperparameters. This reduction changes the sampling space from the sum of parameters and hyperparameters in hierarchical Bayesian inference to the number of parameters, while it provides comparable results to hierarchical Bayesian inference. As the number of questionable parameters increases, the increasing dimensionality of the parameter and hyperparameter space will begin to affect the ability to sufficiently explore the combined parameter and hyperparameter space and generate samples from the posterior effectively, which restricts the use of hierarchical Bayesian inference.
}

\bibliographystyle{unsrt}  
\bibliography{references}

\end{document}